\documentstyle[12pt,cite,epsfig,concrete]{article}
\begin{document}
\baselineskip=16pt

\centerline{\bf Role of Electron--Electron Interactions on Spin Effects in}

\centerline{\bf Electron--Hole Recombination in Organic Light Emitting Diodes.}
\vskip 1.0 truein
\centerline{Abstract}
\vskip 1pc
We extend our theory of electron--hole recombination in organic light emitting 
diodes to investigate the possibility that high energy singlet and triplet 
excited states with large electron--hole separations are generated in
such processes, over and above the lowest singlet and triplet excitons. Our
approach involves a time-dependent calculation of the interchain / 
intermolecular charge--transfer within model Hamiltonians that explicitly
include electron-electron interactions between the $\pi$-electrons. We show 
that the electron--hole recombination reaction can be 
viewed as a tunneling process whose cross section
depends on both the matrix element of the interchain part of Hamiltonian 
and the energy difference between the initial polaron--pair
state and the final neutral states. There occurs a bifurcation 
of the electron--hole recombination path in each of the two
spin channels that leads to the generation of both the lowest energy 
exciton and a specific high energy charge-transfer state, with the matrix 
elements favoring the lowest energy exciton and the energy difference factor 
favoring the higher energy state.  The overall effect of the 
electron--electron interactions is to enhance the
singlet:triplet yield ratio over the value of 0.25 predicted 
from statistical considerations that are valid only within noninteracting 
models.
\pagebreak
\section{\bf Introduction}

The fundamental electron-hole (e-h) recombination process in organic light 
emitting diodes (OLEDs)
can be written as,
\begin{equation}
\label{CT}
P^+ + P^- \to G + S/T
\end{equation}
where $P^{\pm}$ are charged polaronic states of the emissive molecule, $G$
is the ground state of the neutral molecule, 
$S$ is the singlet excited state and $T$ the triplet excited state
of the neutral molecule. Eq.~\ref{CT} indicates that
both singlet and triplet excitons are likely products of the e-h
recombination process.  The quantity that determines the efficiency of 
OLEDs is then the fraction of emissive singlets, $\eta$, that are formed 
in the above charge recombination process. Since electrons and holes are 
injected randomly in the device, only 25\% of the initial polaron-pair 
states $|P^+P^-\rangle$ are singlets. If it is now assumed that the 
rate constant for the e-h recombination in reaction \ref{CT} is equal for 
the singlet and the triplet channels (as would be true in the 
independent-electron limit), one arrives at a theoretical upper bound 
for $\eta$ at 0.25. 
Many experimental studies, however, point to breaching of this upper 
bound in real systems \cite{cao,pk,Wilson01a,tandon,Greenham,
Wohlgenannt02a,Virgili}
and $\eta$ values considerably larger than 0.25 have been claimed.

At the same time though, there exists other experimental work that views these
results with skepticism \cite{Baldo99a,segal,Li04}. The latter authors argue 
that the value of $\eta$ is decided entirely by the fraction of initial bound 
polarons $|P^+P^-\rangle$ that are singlets, i.e., 0.25, and no further change 
in this quantity can take place, irrespective of the rates of the e-h 
recombination rates in the singlet and triplet channels in Eq.~(1).
Theoretical attempts to resolve this paradox recognize the important
role of electron correlations in all cases, but can nevertheless be 
classified into two broad categories.
In one, the focus has mostly been on the {\it lowest}
singlet and triplet excited states, as in the independent
electron model \cite{Shuai00a,shuai2,Tandon,Moushumi}, and the 
the calculation of the relative cross-sections of the e-h recombinations 
within the singlet and triplet channels were carried out in detail.   
Although the actual calculations are quite different
within the different theoretical approaches within this category,
the calculated yield of the lowest singlet exciton is greater than 0.25 
in all cases. Within the second category of theoretical work, it is 
tacitly assumed (but not proved) that (i) the initial products of the e-h 
recombination are {\it higher energy} singlet and triplet states, and (ii) the
yields of these excited singlets and triplets are in the ratio 1:3 (in 
conformity with the independent-electron statistics). Within this category 
of models it is the subsequent relaxations from the high energy states to the
lowest singlet and triplet excitons that gives the change in the ratio 
of emissive singlets to triplets \cite{hong,bittner}. 
As discussed later, from energetic considerations alone there is distinct
possibility that the products of the e-h recombination reaction are the
high energy excited states rather than the lowest excitons. The question 
therefore arises which of these two approaches, if any, describe best 
the e-h recombination in the real materials.

In view of these diverse experimental and theoretical results, we revisit
our original work \cite{Tandon,Moushumi} to investigate the possibility 
that the products of the e-h recombination reaction are high energy 
excited states, and to determine
to what extent the singlet:triplet yield ratio is affected by this. We
show that in both the singlet and triplet channels, there exists strong
likelihood of bifurcation of the reaction paths, with one path leading
to the lowest exciton, the other leading to a specific excited state.
Assuming now the applicability of Kasha's rule \cite{kasha} 
within both spin channels
(i.e., assuming that in both channels, the higher energy excited state decays
to the lowest exciton in ultrafast time scale) one can in principle
estimate the total yields of the lowest singlet and triplet states and
the overall singlet:triplet ratio. This is what is attempted in the
present paper. 

In section 2 we present a brief review of the experimental results. 
In section 3 we present a mechanistic discussion of the e-h recombination 
for noninteracting electrons, with emphasis on the degeneracies that 
characterize this limit and the consequences thereof. We follow this with
a summary on the  the nature of excited states in conjugated polymers in 
section 4.  These results are useful in obtaining physical understanding 
of the numerical results that we obtain for the e-h recombination.
In sections 5 and 6, we present our theoretical model for the
e-h recombination as well as detailed numerical results for various cases.
In the Conclusion section we end with some basic issues that need to 
be addressed in the future and discuss our viewpoints regarding these issues.

\section{\bf Brief survey of experimental results}

The basic difference between electroluminescence (EL) 
and photoluminescence (PL) in molecular materials lies in the 
initial process by which the excited state is formed. Independent of 
this step, in both cases, as per the Kasha rule 
\cite{kasha} fluorescent emission occurs from the lowest excited 
singlet electronic state (with very few exceptions).
This is because rapid internal conversion funnels higher excited 
states to the lowest excited state of the same spin symmetry,
whenever the equilibrium geometries of the initial and final 
excited states are not very different. Thus the cross-section for 
the final process of light emission is nearly the same in both EL and PL. 
However, while the formation of the emissive species, the singlet 
optical exciton, has a quantum efficiency (QE) of nearly 1 in the 
PL process, this QE is $\eta$ in the case of EL.  Hence, the ratio 
of the EL to PL efficiency provides a lower bound on $\eta$ in the 
e-h recombination in the EL process. Using this principle,
experimentally $\eta$ has been found to range from $\approx$ 0.25 to 0.66
in different materials \cite{cao,pk,tandon}. In OLEDs containing molecular 
components as the emissive species, such  as aluminum tris (8-hydroxy 
quinoline) (Alq$_3$), $\eta$ has been determined to be close to the  
independent-electron statistical value \cite{Baldo99a,segal}. On the 
other hand, considerably larger $\eta$ of 0.45 has been claimed in 
derivatives of poly (para phenylene vinylene)
(PPV) by Cao {\it et al.}\cite{cao} and Ho {\it et al.} \cite{pk}.
$\eta$ has also been measured by photoinduced absorption 
detected magnetic resonance (PADMR) \cite{tandon,Wohlgenannt02a}. 
In this technique a magnetic field of about 0.1 Tesla 
is applied to Zeeman split the spin-1/2 states of the charged 
polarons at a temperature of about 20 $K$. Application
of an intense microwave to match the Zeeman splitting leads to equal 
populations of the up and down spin states. This in turn would lead to 
higher probability of two neighboring polarons of opposite charge 
having antiparallel spin orientations than parallel spin orientations. 
Thus, the recombination will yield fewer triplets than in the absence 
of a field and a saturating microwave, resulting in attenuation 
of the triplet absorption in the PADMR experiment. From measurement 
of this attenuation, it is possible to calculate $\eta$.  Wohlgenannt 
{\it et al.} \cite{tandon,Wohlgenannt02a} have determined $\eta$ 
for a large number of polymeric materials this way and have found it 
to be strongly material dependent;  in all the cases they studied, 
$\eta$ was determined to be larger than 0.25.  More recently, Wilson
{\it et al.} \cite{Wilson01a} and Wohlgenannt {\it et al.}
\cite{Wohlgenannt02a} have shown that $\eta$ varies with conjugation 
length, from 0.25 for small molecules to considerably larger for long 
chain oligomers.  These results are in contradiction with the claim 
in reference \cite{segal} that in electroluminescent devices with 
MEH-PPV as the emissive material the singlet fraction is (20$\pm$4)\%. 

Increase in the population of polaron pairs with antiparallel spins 
under resonance condition also implies formation of a higher fraction 
of singlet excitons. Monitoring EL intensity under 
resonance conditions (ELDMR) should give an enhancement consistent 
with the decrease in PA intensity under the same conditions. This has
indeed been observed by Segal {\it et al.} \cite{segal} as well as Li 
{\it et al.} \cite{Li04} in Alq$_3$. 
Since estimates based on other measurements give $\eta \sim 0.25$ 
for Alq$_3$ \cite{Baldo99a,segal}, Li {\it et al.} attribute this 
increase in EL to reduced polaron population under resonance leading 
to reduced quenching of the singlet excitons \cite{Li04}.
Reduced polaron population results, however, from increased recombination 
process and hence increased population of singlet excitons.  Thus in 
polymeric materials, the enhanced ELDMR could be both due to higher 
rate of recombination process and reduced rate of quenching of the 
singlet excitons.

Finally a recent experimental paper by Lin {\it et al.} has claimed 
that $\eta$ can be {\it smaller} than 0.25 in LED devices under low 
electric field \cite{lin}.  This work, however, has been severely 
criticized by authors who have pointed out that the experiments were 
carried out at room temperature, and the absorption spectrum assigned 
to triplets by Lin {\it et al.} was actually due to charged polarons 
\cite{Osterbacka,Dhoot}.  Furthermore, the critical assumption in the 
model, namely, that no triplet states exist between the conduction band 
edge triplet and the lowest triplet is not borne by the exact triplet 
spectrum calculated for long conjugated chains \cite{MDas} using the 
widely accepted Pariser--Parr--Pople model \cite{PPP}.

\section{Two-level picture of e-h recombination}

In view of what follows, we discuss here briefly the recombination 
reaction (1) for independent electrons (H\"uckel model). The electronic 
structures of all the components in Eq.~(1) are given by single molecular 
configurations in this limit, and for arbitrary chain lengths or molecular 
sizes, only the highest occupied and lowest unoccupied molecular orbitals 
(HOMO and LUMO) are relevant. Each component of reaction (1) can therefore 
be described within a two-level scheme, as shown in Fig.~1. The total 
energy of the initial state $|P^+ \cdot P^-\rangle$ and the final state 
$|G \cdot S/T\rangle$ in Fig.~1 are identical. Hence if we view the
e-h recombination as a tunneling process, there occurs resonant 
tunneling from the initial polaronic states to the final neutral 
states, if the matrix element of the perturbation (corresponding to 
the transfer term which causes an electron to hop between the 
conjugated chains) connecting the initial and final states are nonzero. 
The matrix elements $\langle G \cdot S|H|P^+P^-\rangle_S$ and
$\langle G \cdot T|H|P^+P^-\rangle_T$, where $| P^+P^-\rangle_S$ and 
$| P^+P^-\rangle_T$ are singlet and triplet polaron-pair species,
are identical for arbitrary interchain hopping and hence the tunneling 
probability is the same in the singlet and the three triplet channels. 
This quantum resonant tunneling picture leads us to $\eta~=~0.25$. 
Notice that this requires the strict degeneracy
\begin{equation}
\label{degeneracy}
E(P^+) + E(P^-) = E(G) + E(S/T)
\end{equation}
where the energy E in each case refers to the total energy of the state 
in question.  If for any reason these equalities are not obeyed, in 
particular if $E(S) \neq E(T)$ there is no reason to have $\eta$ = 0.25.
~\\

\begin{figure}[h]
\begin{center}
{\includegraphics[width=10cm,height=3cm]{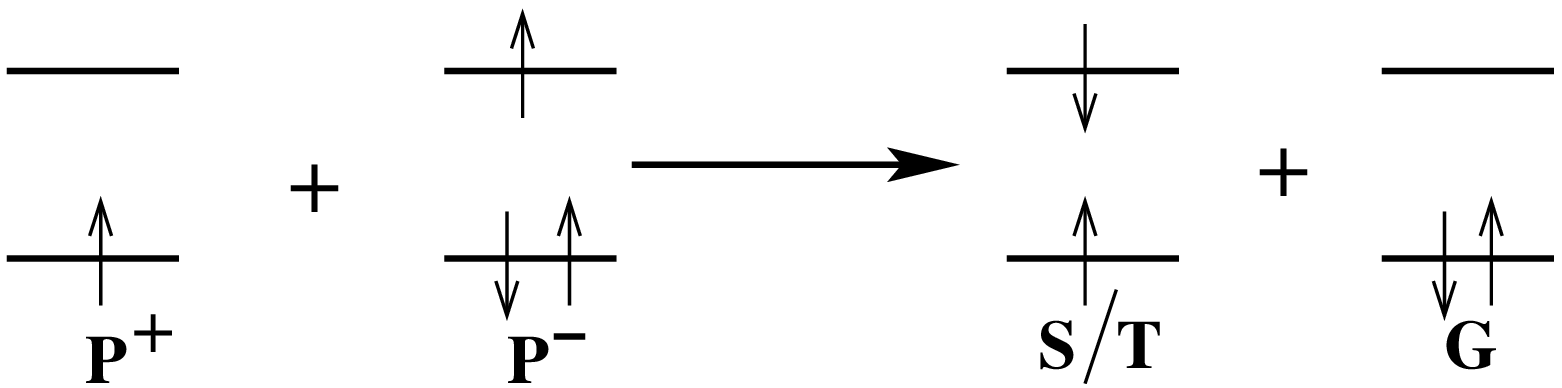}}
\end{center}
\end{figure}
\noindent
FIG. 1.{\it{The HOMO and LUMO orbital occupancies for the initial
polaronic pair state and the final neutral states for the total
z-component of the spin $M_s$=0.}}

\section{\bf Excited states in conjugated polymers}

We briefly review here the nature of the excited states in conjugated
polymers within correlated electron models. This will be useful in 
understanding the bifurcation of the e-h recombination paths (1) mentioned
in the above.

The noninteracting electrons description of conjugated polymers 
is based on the $\pi$-electron H\"uckel model. The H\"uckel molecular 
orbitals (HMOs) are obtained as  linear combinations of the atomic 
$2p_z$ orbitals, one at each carbon atom in conjugation. Usual quantum 
chemical approaches that go beyond H\"uckel theory use the HMOs as the 
starting point and include electron correlations via a configuration 
interaction (CI) scheme by using a restricted number of excited HMO 
configurations such as the singly excited and the doubly excited 
configurations in the singles and doubles CI (SDCI) approach. However, when 
the strength of repulsion between two electrons occupying the same 
$2p_z$ orbital is comparable to the energy difference between the LUMO and
the HOMO, it is preferable to start with the electron configurations in 
the atomic orbital (AO) basis.  This is particularly important in the 
polymer limit since approaches such as SDCI are not size-consistent and 
size-consistent techniques such as perturbation methods do not converge 
in the regime of intermediate correlations. The guiding factors in this 
regime would be the physical insights developed from a real space or AO 
picture.  In this section we will illustrate how this picture helps in 
understanding the excitations of a conjugated polymer chain.

\begin{figure}
\begin{center}
{\includegraphics[width=12cm,height=7cm]{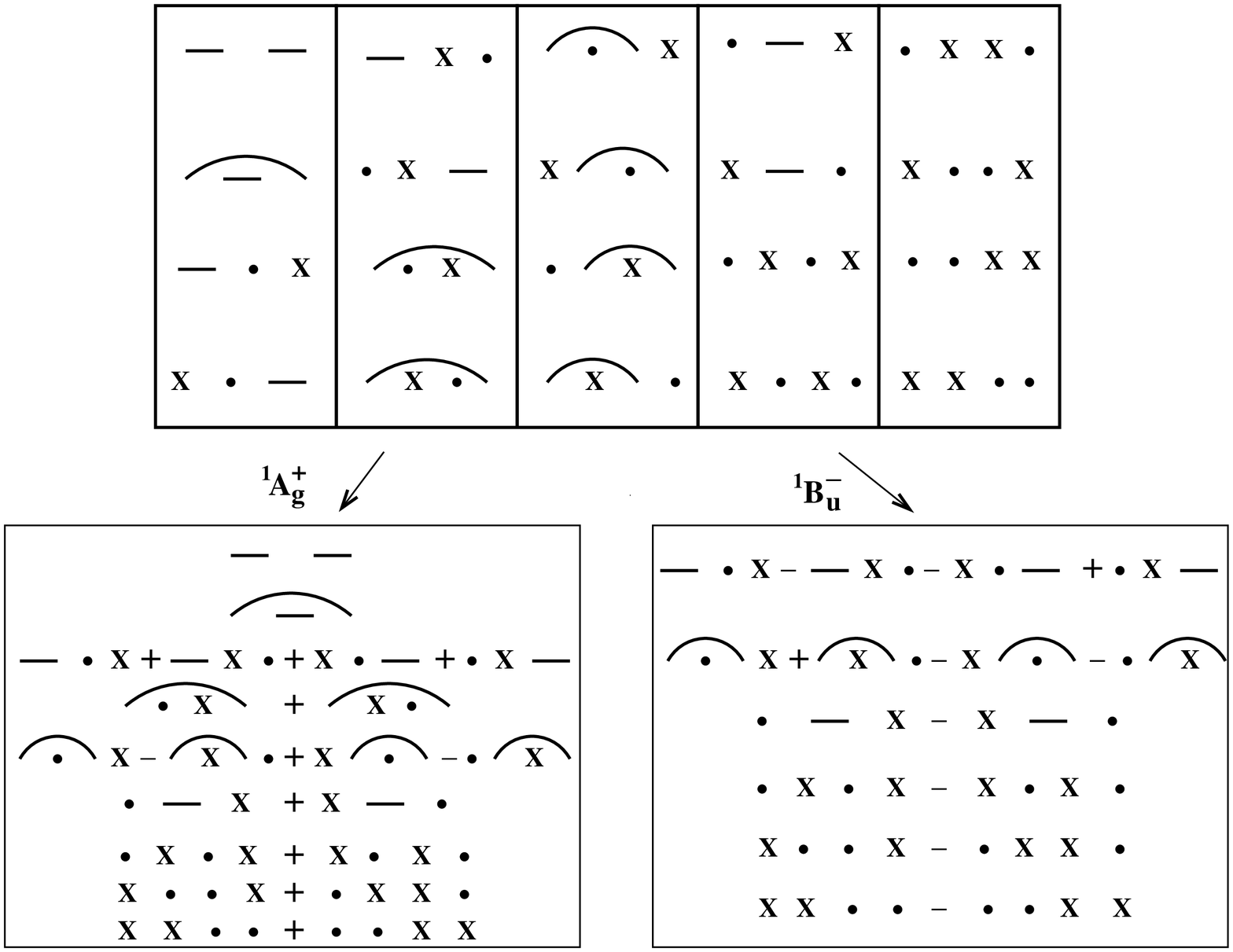}}
\begin{center}
(a)
\end{center}
\vspace*{0.5cm}
{\includegraphics[width=12cm,height=7cm]{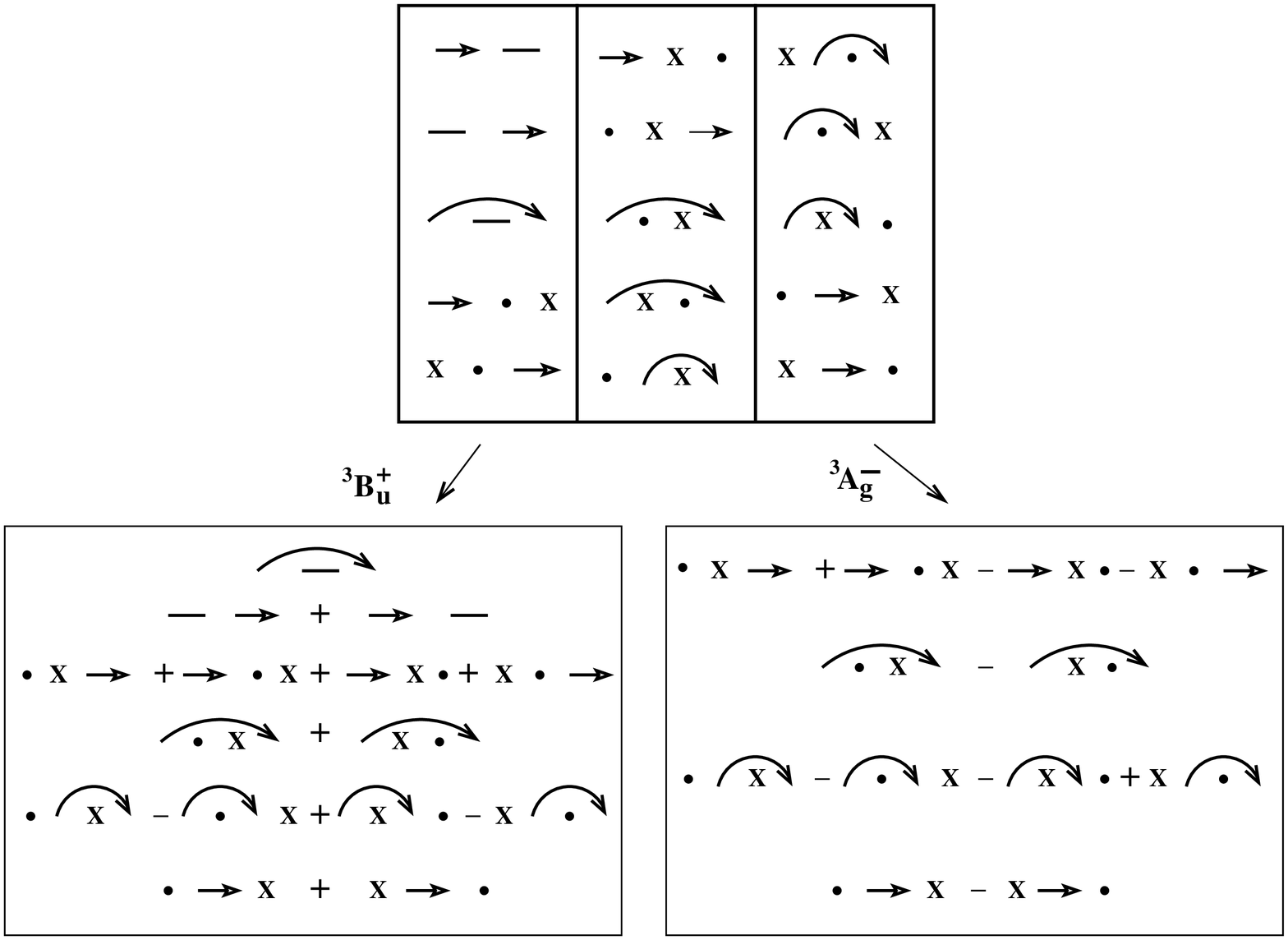}}
\begin{center}
(b)
\end{center}
\end{center}
FIG. 2. {\it The VB basis states for singlets and triplets for butadiene.
Each cross (X) represents $2p_z$ orbital at the site occupied by two 
electrons, a dot ($\cdot$) represents an empty orbital and a line (arrow) 
between two sites represents singlet (triplet) spin-pairing of the singly 
occupied $2p_z$ orbitals at the sites. In (a) the twenty singlet VB diagrams 
yielding nine symmetrized singlet basis in the $^1A_g^+$ and six symmetrized 
singlet basis in the $^1B_u^-$ subspaces are shown. The ground state lies 
in the $^1A_g^+$ subspace while the optically allowed excitations from 
the ground state lie in the $^1B_u^-$ subspace. Two other subspaces 
corresponding to $^1A_g^-$ and $^1B_u^+$ subspaces are not shown. In (b) 
the fifteen triplet VB diagrams, the six basis states in the $^3B_u^+$ 
subspace, and the four basis states in the $^3A_g^-$ subspace are shown.
The lowest triplet lies in the $^3B_u^+$ subspace.} 
\end{figure}

We begin with the analysis of the simple case of butadiene with
four $2p_z$ orbitals (N=4) in conjugation. The number of electrons 
N$_e$ occupying the N orbitals is also 4.  The models we deal with 
are nonrelativistic (since spin-orbit interactions are negligible) 
and hence total spin, $S$, as well as the z-component of total spin, 
$M_S$, are well defined.  We can write 36 distinct (linearly independent) 
electron configurations corresponding to $M_s=0$ for the case of N$_e$ 
= N = 4.  Constructing basis functions with fixed total spin, such 
as S = 0 singlet basis functions or S = 1 triplet basis functions is 
nontrivial. The approach that has proved to be useful and physically 
appealing has been to use valence bond (VB) functions \cite{Ramasesha84a}, 
which are linear combinations of the Slater determinants corresponding to 
different MO configurations. The VB basis states are completely 
defined if (i) the orbital occupancies by electrons
are defined and (ii) the spin-pairing among the 
singly occupied orbitals are provided. VB singlets
are represented by lines connecting the orbital pair 
(see Fig. 2 and discussion below). For VB basis with $S \neq 0$
and $M_S=S$,
besides the singlet pairings, it is also necessary to specify 
the $2S$ AOs with parallel spin occupancies. Within our VB theory such
sites are connected by a triplet bond, represented by arrows (see Fig. 2).

In Figs. 2 (a) and (b) we show the twenty singlet and fifteen 
triplet VB diagrams for $N=N_e=4$.  Linear chains have mirror 
plane as well as inversion symmetries, implying that all basis 
functions as well as eigenstates can be classified as having even 
spatial parity ($A_g$) and odd spatial parity($B_u$). The bipartite 
nature of linear polyenes (as well as most conjugated polymers) 
also implies charge-conjugation symmetry; thus each spatial symmetry 
subspace can be further partitioned into even and odd charge-conjugation 
symmetries, giving four different symmetry subspaces overall. We have 
given in Fig.~2(a) the VB basis functions, -- superpositions of VB 
diagrams, -- that form the  S = 0 $A_g^+$ and $B_u^-$ subspaces 
(the other two subspaces, $A_g^-$ and $B_u^+$, are not shown). 
Similarly in Fig.~2(b) we have shown the $B_u^+$ and $A_g^-$
S = 1 basis functions.

Broadly, the basis functions in Figs.~2(a) and (b) can be classified as 
covalent, i.e., consisting of VB diagrams in which all atoms are 
singly occupied (the first two basis functions in the $^1A_g^+$
subspace, as well as the first two basis functions in the $^3B_u^+$ 
subspace), and ionic, with at least one doubly occupied site and one 
empty site (all other basis functions in Figs.~2). The ionic basis 
functions can be further classified into singly ionic (with one doubly 
occupied and one empty site), doubly ionic (with two doubly occupied and 
two empty sites), and so on (higher ionicities occurring in N$_e$=N $>$ 4). 
The ground state lies in the $A_g^+$ subspace (and henceforth is referred 
as the 1$^1A_g^+$), and with increasing electron-electron interactions, 
the ionicity of the ground state decreases and the wavefunction is more 
strongly dominated by the covalent basis functions (since the on-site part 
of the electron correlations raise the energies of basis functions with 
double occupancies).  Optical excitation from this state is to the lowest 
$^1B_u^-$ state (hereafter 1$^1B_u^-$), which, as seen from Fig.~2(a), 
is necessarily ionic as there exist no covalent VB diagrams in the 
$^1B_u^-$ subspace. In contrast, the lowest triplet states in the 
$^3B_u^+$ subspace again have strong covalent contributions, and hence
are lower in energy than the 1$^1B_u^-$. There exist a ``band'' of such 
triplet states between the 1$^1A_g^+$ and the 1$^1B_u^-$ in the long 
chain limit.
The 1$^3B_u^+$ is higher in energy than the 1$^1A_g^+$, since while there 
can be charge-transfer delocalization across an arbitrary singlet
bond, there is no such charge-transfer across triplet bonds.
From the physical natures of the basis functions then, it is clear that several 
of the lowest triplets will occur below the 1$^1B_u^-$. 

If the electron correlations include intersite Coulomb interactions over 
and above on-site correlations, singly ionic VB basis functions with 
neighboring double occupancy (particle) and vacancy (hole) (the third 
basis function in the $^1A_g^+$ subspace and the first basis function in
the $^1B_u^-$ subspace) have lower energy than basis functions in which the 
double occupancy and the vacancy are further away. In the long chain limit 
there are many more (practically infinite) basis functions that belong to 
the latter class and few with short range separations between the double 
occupancy and the vacancy.  We see therefore that for realistic strong 
Coulomb interactions that characterize conjugated polymers there will be 
a strong tendency for formation of excitons, with a few of the states 
dominated by ionic VB functions with small separations between the double 
occupancy and the vacancy splitting off from the continuum of singly ionic 
states. 

The above conjectures based on the physical nature of the VB basis functions
have been substantiated by numerical calculations by several groups
\cite{DGuo,Abe,Barford,adqc} and is also supported by experiments 
\cite{DGuo,Weiser,Leng,Liess,Blatchford,Sinclair,Samuel,Leng1,Hsu,Yan}.
In Fig.~3 we have shown the schematic energy spectrum of conjugated 
polymers. The singlet and triplet exciton states below the conduction 
band edge that are labeled as m$^1A^+_g$ and m$^3A^-_g$ have been 
discussed extensively in the context of optical nonlinearity and are 
characterized by very large transition dipole couplings to the 1$^1B^-_u$ 
and 1$^3B^+_u$ states respectively \cite{DGuo,Abe,Barford,adqc}.  
Energetically, the m$^1A^+_g$ is degenerate with the m$^3A^-_g$ at 
the level of singles-CI \cite{Abe} and very slightly above the m$^3A^-_g$ 
within exact calculations \cite{Shimoi}. Still higher energy singlet and
triplet states with large transition dipoles to the m$^1A^+_g$ and
m$^3A^-_g$ states are the n$^1B^-_u$ and n$^3B^+_u$
\cite{DGuo,Abe,Barford}, also included in Fig.~3. The n$^1B_u^-$ 
lies at the edge of the continuum threshold, as has been shown from 
earlier calculations \cite{DGuo,Abe,Barford}. The n$^3B^+_u$ has not 
been discussed previously. We have calculated the energy
of this state exactly for a large range of parameters (see below) 
and have found in all cases this state to be nearly degenerate with 
the m$^3A^-_g$ and invariably below the n$^1B_u^-$. Although the 
bulk of the energetics calculations are for linear chain polyacetylenes 
and polydiacetylenes  \cite{DGuo,Abe,Barford}, work by different groups 
have indicated that the same basic energy level scheme applies also to 
conjugated polymers with aromatic groups, in the energy region up to and 
including the conduction band threshold \cite{soosprl,Chandross,shuai97}.
It is natural to assume that the ground state of the polaron pair formed 
in the OLEDs, $|P^+P^-\rangle$ is just at the bottom of the conduction 
band edge. This state is also included in Fig.~3, where we have
made no distinction between $|P^+P^-\rangle_S$ and $|P^+P^-\rangle_T$, as the
energy difference between the singlet and triplet polaron-pair states is tiny
\cite{Kadashchuk} and would be invisible on the scale of Fig.~3.
~\\
\begin{center}
{\includegraphics[width=7cm,height=7cm]{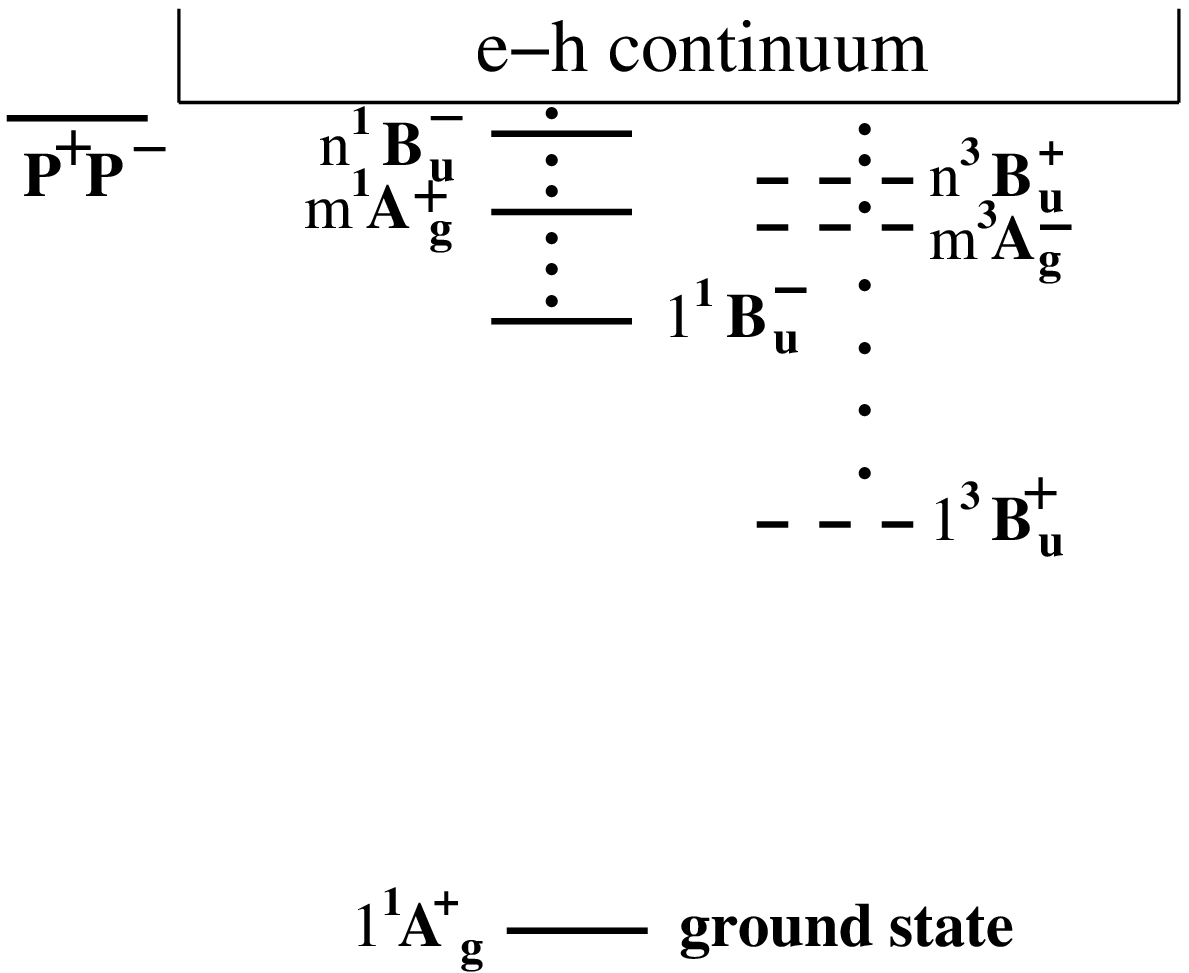}}
\end{center}
{FIG. 3.} {\it Schematic excitation spectrum of a conjugated polymer. 
Dots indicate that there are several excitations between the 
labeled states.}

\section{Correlated electron theory of e-h recombination}

Our goal in this section is to lay out the formalism for the detailed 
calculations of relative formation rates of singlet and triplet excitons 
starting from an oppositely charged polaron pair (see reaction (1)), 
in view of the correlated electron description of the electronic 
structure of $\pi$-conjugated polymers (see Fig.~3). In contrast 
to our earlier work \cite{Tandon,Moushumi}, we recognize at the outset 
that e-h recombination can generate $S$ and $T$ states at energies 
higher than the lowest energy excitons.  We calculate the yields to 
all such states under different conditions using the techniques 
developed in \cite{Tandon,Moushumi}.  Our procedure consists in 
computing the time-dependent evolution of oppositely charged polyene 
molecules under the influence of the composite two-chain Hamiltonian, 
as discussed below.  As we limit ourselves to calculations based on 
the rigid bond approximation and small finite molecules, we assume 
that all high energy singlets and triplets decay in ultrafast times 
to the 1$^1B_u^-$ and 1$^3B_u^+$, respectively. We believe that this 
assumption is valid for the real systems \cite{MDas}.

\subsection{The model system}

Our model system consists of two polyene chains of equal lengths
that lie directly on top of each other, separated by 4 \AA. 
We consider the charge recombination process of Eq.~\ref{CT}, and there
are two possible initial states: (i) a specific chain (say chain 1)
is positively charged,
with the other (chain 2) having negative charge, a configuration that 
hereafter we denote as $P^+_1P^-_2$, where the subscripts 
1 and 2 are chain indices, or (ii) 
the superposition $P^+_1P^-_2 \pm P^+_2P^-_1$, in the same notation. In our 
calculations we have chosen the first as the proper initial state, since
experimentally in the OLEDS the symmetry between the chains
is broken by the external electric
field (we emphasize that the consequence of choosing the symmetric or 
antisymmetric superposition can be easily predicted from all our 
numerical calculations that follow). Even with initial state (i),
the final state can consist of both 
$G_1 \cdot S_2$ and
$G_2 \cdot S_1$ in the singlet channel. The same is true in the
triplet channel, i.e., either of the two chains can be in the ground 
(excited) state.
Hereafter we will write the initial states as $|i_S\rangle$ and
$|i_T\rangle$, where the subscripts $S$ and $T$ correspond to spin states
$S$ = 0 and 1. We consider only the $M_S$ = 0 triplet state.
In the absence of an external magnetic field the e-h recombination
reaction rate for all three triplet channels with different $M_S$ are the same.
The initial states are simply the product states with 
appropriate spin combinations,

\begin{eqnarray}
\label{initial_states}
|i_S\rangle = 2^{-1/2}(|P^+_{1,\uparrow}\rangle |P^-_{2,\downarrow}\rangle -
|P^+_{1,\downarrow}\rangle |P^-_{2,\uparrow}\rangle) \\
 \nonumber \\
|i_T\rangle = 2^{-1/2}(|P^+_{1,\uparrow}\rangle |P^-_{2,\downarrow}\rangle +
|P^+_{1,\downarrow}\rangle |P^-_{2,\uparrow}\rangle)
\end{eqnarray}

The overall Hamiltonian for our composite two-chain system consists
of an intrachain part $H_{intra}$ and an interchain part $H_{inter}$. 
$H_{intra}$ describes individual 
chains within the Pariser-Parr-Pople (PPP) Hamiltonian \cite{PPP} for $\pi$-electron
systems, written as,

\begin{eqnarray}
H_{intra} = -\sum_{<ij>,\sigma}t_{ij} (a^{\dagger}_{i,\sigma}
a^{}_{j,\sigma} + H.C.) + \sum_i \epsilon_i n_i + \nonumber \\
\sum_i U_i n_{i,\uparrow}n_{i,\downarrow} 
+  \sum _{i>j} V_{ij} (n_i -z_i) (n_j -z_j)  
\label{PPP}
\end{eqnarray}

\noindent where $a^{\dagger}_{i,\sigma}$ creates a $\pi$-electron of 
spin $\sigma$ on
carbon atom $i$, $n_{i,\sigma} = a^{\dagger}_{i,\sigma}a^{}_{i,\sigma}$ is 
the number of electrons on atom $i$ with spin $\sigma$ and 
$n_i = \sum_{\sigma}n_{i,\sigma}$ is the total number of electrons on atom 
$i$, $\epsilon_i$ is the site energy and $z_i$ are the local chemical 
potentials. The hopping matrix element 
$t_{ij}$ in the above are restricted to nearest neighbors and 
in principle can contain electron-phonon interactions, although 
a rigid bond approximation is used here. $U_i$ and $V_{ij}$ are the
on-site and intrachain intersite Coulomb interactions. 

We use standard parameterizations for $H_{intra}$. The hopping integrals 
for single and double bonds are taken to be 2.232 eV and 2.568 eV, 
respectively and all site energies, for simple polyenes 
with all sites equivalent, are set to zero. We choose the Hubbard 
interaction parameter $U_C$ for carbon to be 11.26 eV, and for the 
$V_{ij}$ we choose the Ohno parameterization \cite{Ohno64},
\begin{eqnarray}
V_{ij} = 14.397\left[\left({{28.794}\over {U_i+U_j}}\right)^2~+~r^2_{ij}
\right]^{-{{1} \over {2}}}
\label{Ohno}
\end{eqnarray}
where the distance $r_{ij}$ is in \AA, ~$V_{ij}$ is in eV and the local 
chemical potential $z_C$ for $sp^2$ carbon is one. It should be noted then
when heteroatoms like nitrogen are present, the on-site correlation energy,
the site energy and the local chemical potential could be different from 
those for carbon \cite{Tandon}. 
For $H_{inter}$, we choose the following form,
\begin{eqnarray}
\label{inter}
H_{inter} = -t_{\perp}\sum_{i,\sigma}(a^{\dagger}_{i \sigma}
a^{\prime}_{i,\sigma} + H.C.) + \nonumber \\
+ X_{\perp}\sum_{i,\sigma}(n_i + n^{\prime}_i)(a^{\dagger}_{i \sigma}
a^{\prime}_{i,\sigma} + H.C.) + \nonumber \\
\sum_{i,j} V_{i,j}(n_i -z_i)(n^{\prime}_j - z_{j^\prime})
\end{eqnarray}
In the above, primed and unprimed operators 
refer to corresponding
sites on different 
chains. Note that the interchain hopping $t_{\perp}$ is restricted to 
nearest interchain neighbors.
The interchain Coulomb interaction $V_{i,j}$, however, includes interaction 
between any site on one chain with any other site on the other chain.
In addition to the usual one-electron hopping that occurs within the zero 
differential overlap approximation we have also included a many-electron 
site charge-bond charge repulsion $X_{\perp}$ (operating between nearest 
interchain neighbors only) that consists of multicenter Coulomb integrals
\cite{PPP,Campbell90a}. 
This term should also occur within $H_{intra}$, but is usually ignored 
there because of its small magnitude, relative to all other terms. In 
contrast, the $t_{\perp}$ in $H_{inter}$ is expected to be much smaller, 
and $X_{\perp}$ cannot be ignored in interchain processes, especially at 
large interchain separations \cite{Rice96a}. 
We have done calculations for both $X_{\perp}$
= 0 and $X_{\perp} \neq$ 0.

\subsection{Time-evolution of the polaron pair state}

Our approach consists in calculating the time-evolution of the
initial states $|i_S\rangle$ and $|i_T\rangle$ [see Eqs. (3) and (4)]
under the influence of the full Hamiltonian, and then evaluating 
the overlaps of the time-evolved states
with all possible final states $|f_S\rangle$ and $|f_T\rangle$ of the 
individual neutral chains. In OLEDs, the
$P^{\pm}$ are created at opposite ends of the device and they execute hopping
motion towards each other under the influence of an external electric field
($P^{\pm} + G \to G + P^{\pm}$). The polaron wavefunctions remain unperturbed 
throughout this process, until they are within the radius of influence of each 
other. We define this particular instant as time $t$ = 0, and we visualize 
that the interchain interactions $H_{inter}$ are ``switched on'' suddenly 
from zero at this time. The intermolecular charge-transfer (CT) hereafter 
is rapid (a few to several tens of femtoseconds, for realistic interchain 
hopping $t_{\perp}$, see below). It is the ultrashort timescale of this CT 
process that justifies the choice of our initial state. 

In principle, given a Hamiltonian, propagation of any initial state is easily 
achieved by solving the
time-dependent Schrodinger equation. One could use the interaction picture
to separate the nontrivial evolution of the initial state from the trivial
component which occurs as a result of the evolution of the product of the
eigenstates of the Hamiltonian of the subsystems. In the context of the 
many-body
PPP Hamiltonian such an approach is difficult to implement numerically. This
is because the total number of eigenstates for the two-chain system is very
large: the number of such states for two chains of six carbon atoms each is
853,776 in the $M_s$ = 0 subspace. Obtaining all the eigenstates of the
two-component system and expressing the matrix elements of $H_{inter}$ in the
basis of these eigenstates is therefore very intensive computationally. It is
simpler to calculate the time evolution in the Schrodinger representation,
determine the time-evolved states, and project them on to the desired
final eigenstates (for instance, $|1^1A_g\rangle_1|1^1B_u\rangle_2$). This 
is the approach we take.

Henceforth we refer to the initial states  $|i_S\rangle$ and $|i_T\rangle$ 
collectively as $\Psi(0)$ and the time-evolved states as $\Psi(t)$. In 
principle, the time evolution can be done by operating on $\Psi(0)$ with 
the time evolution operator,
\begin{equation}
U(0,t) = \exp(-iHt)
\label{time}
\end{equation}
where $H$ is the total Hamiltonian $H_{intra} + H_{inter}$. This approach 
would, however, require obtaining a matrix representation of the exponential
time evolution operator, which again requires the determination of the
prohibitively large number of eigenstates of the composite two-chain system.
We can avoid this problem by using small discrete time intervals and expanding 
the exponential operator in a Taylor series, and stopping at the linear term.
Such an approach, however, has the undesirable effect of spoiling unitarity, 
and for long time evolutions would lead to loss of normalization of the 
evolved state.
The way around this dilemma has been proposed and used by others
\cite{Crank47a,Varga62a} in different contexts and involves using the 
following truncated time-evolution scheme,
\begin{eqnarray}
(1 + iH{{\Delta t} \over {2}}) \Psi(t+\Delta t) =
(1 - iH{{\Delta t} \over {2}}) \Psi(t)
\label{evolve}
\end{eqnarray}
In the above equation, on the left hand side, we evolve the state at time 
$(t+\Delta t)$ backward by $\Delta t/2$ while on the right hand side,
we evolve the state at time $t$ forward by $\Delta t/2$. By forcing these
two to be equal, we ensure unitarity in the time evolution of the state.
It can be seen easily that this time evolution which is accurate to 
${{\Delta t^2} \over {2}}$ is unitary. For a given many-body Hamiltonian 
and initial state, the right hand side of Eq.~\ref{evolve} is a known vector 
in the Hilbert space of the two-chain Hamiltonian.  The left hand side 
corresponds to the action of a matrix on an as yet unknown vector, that 
is obtained by solving the above set of linear algebraic equations. 

After each evolution step, the evolved state is projected onto the space of
neutral product eigenstates of the two-chain system. The relative yield
$I_{mn}(t)$ for a given product state $|m,n\rangle$ = $|m\rangle_1|n\rangle_2$
is then obtained from,
\begin{equation}
I_{mn}(t) = |\langle\Psi(t)|m,n \rangle|^{2}
\label{overlap}
\end{equation}
In our case the states $|m,n\rangle$ can be any of the 
final states of interest,
viz., $|(1^1A_g)_1(1^1B_u)_2\rangle$, $|(1^1A_g)_1(1^3B_u)_2\rangle$, etc.
It is for efficient calculations of the overlaps (while at the same time
maintaining spin purity) in Eq.~\ref{overlap} that we 
transfer
our exact eigenstates of the neutral system in the VB basis to the total 
$M_S$ basis. We emphasize that $I_{mn}(t)$ is a measure of the yield of the 
state $|m,n\rangle$ at time $t$ and is not a cross-section.

\section{Numerical Results}
\label{results}

We first present the results of our calculations of recombination 
dynamics for pairs of ethylenes, butadienes and hexatrienes within 
the noninteracting H\"uckel model ($U_i = V_{ij} = X_{\perp} = 0$) and 
the interacting PPP model. These results have already been discussed 
in detail in reference \cite{tandon}, and hence our presentation 
will be brief. Our calculations here clearly indicate that the yield of 
the lowest singlet exciton is considerably larger than that of the 
lowest triplet exciton within the PPP Hamiltonian. As already discussed 
in section I, however, it is not  necessary that the reaction 
products of the e-h recombination reaction are limited to the lowest 
excitations when e-e interactions are nonzero. A thorough search within 
the PPP model has, however, not detected significant yield of any other 
excited state \cite{tandon}. We discuss why this might be a consequence 
of the small sizes of our model systems. We then present new results of 
calculations 
with $H_{intra}$ as the simple Hubbard Hamiltonian ($V_{ij} = 0$) and 
an extended Hubbard Hamiltonian with short-range intersite Coulomb 
interactions. In spite of the obvious limitations of our finite size
calculations, a mechanism of e-h recombination appears to emerge from 
our work. We are able to show that in both singlet and triplet channels, 
there occurs a bifurcation of the e-h recombination paths, 
with one branch leading 
to the lowest 1$^1B_u$ (1$^3B_u$) exciton, the other leading to the 
formation of the n$^1B_u$ (n$^3B_u$) higher energy CT states. The overall 
S:T ratio then depends on the relative weights of the two branches in each
spin channel, but the fractional singlet exciton yield 
continues to be greater than 0.25.

\subsection{Dynamics in the H\"uckel Model}

While there is no difference in energy between singlets and triplets 
in the H\"uckel Model, it is nevertheless possible to have spin singlet 
and triplet initial states $|i_S\rangle$ and $|i_T\rangle$, as well as
singlet and triplet final states. In Fig. 4 we show the yield for the
e-h recombination in the singlet channel, for pairs of ethylenes, 
butadienes and hexatrienes. The yields for the triplet channels are not 
shown separately in this case, -- we have ascertained that these are 
identical to those in the singlet channel in this case, as expected.
\begin{figure}
\begin{center}
\centerline{\includegraphics[width=10cm,height=6cm]{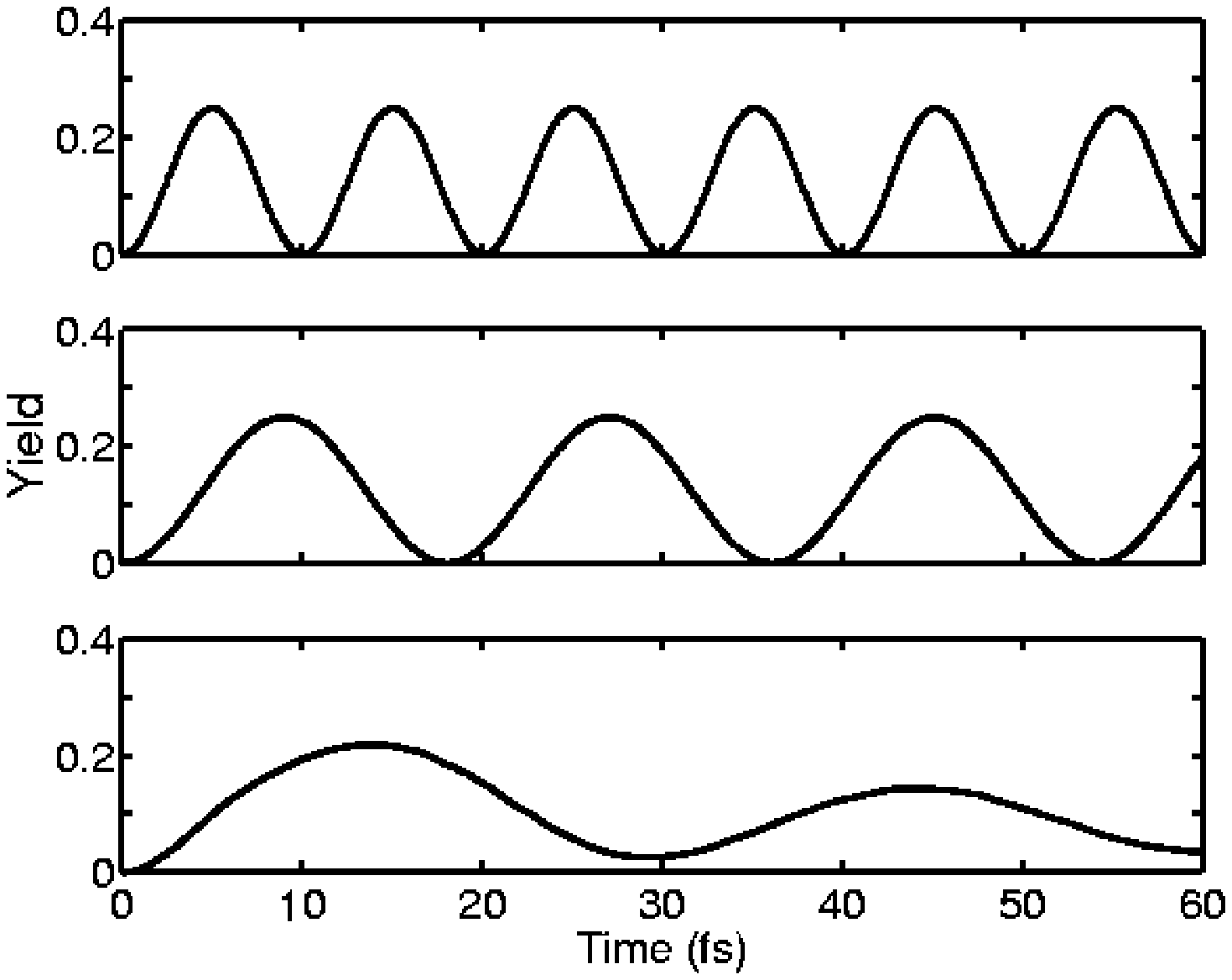}}
\end{center}
{FIG. 4.} {\it Yield in the singlet channel as a function of time, for
pairs of ethylenes (top panel), butadienes (middle panel), and
hexatrienes (bottom panel), within the 
H\"uckel model ($U=V_{ij}=X_{\perp}=0$).}
\end{figure}
These calculations are for $t_{\perp} = 0.1$ eV, $X_{\perp} = 0$ 
within Eq.~\ref{inter}.  
We note that the yields $I_{mn}(t)$ oscillate with time.
This is to be expected within our purely electronic Hamiltonian, within
which an electron or hole jumps back and forth between the two molecular
species. These oscillations are the analogs of the Rabi oscillations 
\cite{Rabi37a,Allen87a} that occur upon the stimulation of a system with 
light, where absorption of light can occur only with nonzero damping. 
Within our purely electronic Hamiltonian, complete transition to the 
final states can only occur in the presence of damping (for example, 
radiative and nonradiative relaxations of the final states), that
has not been explicitly included in our Hamiltonian. The frequency of
oscillation is higher for larger intermolecular transfer integral
$t_{\perp}$, as expected. The frequency of the oscillation also depends
upon the size of the molecule and is lower for larger molecules.
The equalities in the yields of the singlet and triplet excited states
found numerically conforms to the simple free spin statistics 
prediction that the probability of singlet and triplet formation are equal
in the e-h recombination process with $M_s=0$ as the initial state. 
Since the $M_s=\pm 1$ cases always yield triplets, 
the spin statistics corresponding 
to 25\% singlets and 75\% triplets is vindicated in this case.

\subsection{Dynamics in the PPP model}

The results presented in this subsection are for interchain $V_{i,j}$ 
calculated using the Ohno parameters, and interchain hopping $t_{\perp}$ 
= 0.1 eV. We present the results for both $X_{\perp}$ = 0 and 0.1 eV. The top 
left and top right plots in Fig. 5 show the yield $I_{mn}(t)$ in the 
singlet and triplet channels for pairs of ethylenes, butadienes and
hexatrienes, respectively, for the case of $X_{\perp}$ = 0. The same
results are shown in bottom left and bottom right plots for $X_{\perp}$ 
= 0.1 eV.
\begin{center}
\includegraphics[width=6.8cm,height=5.9cm]{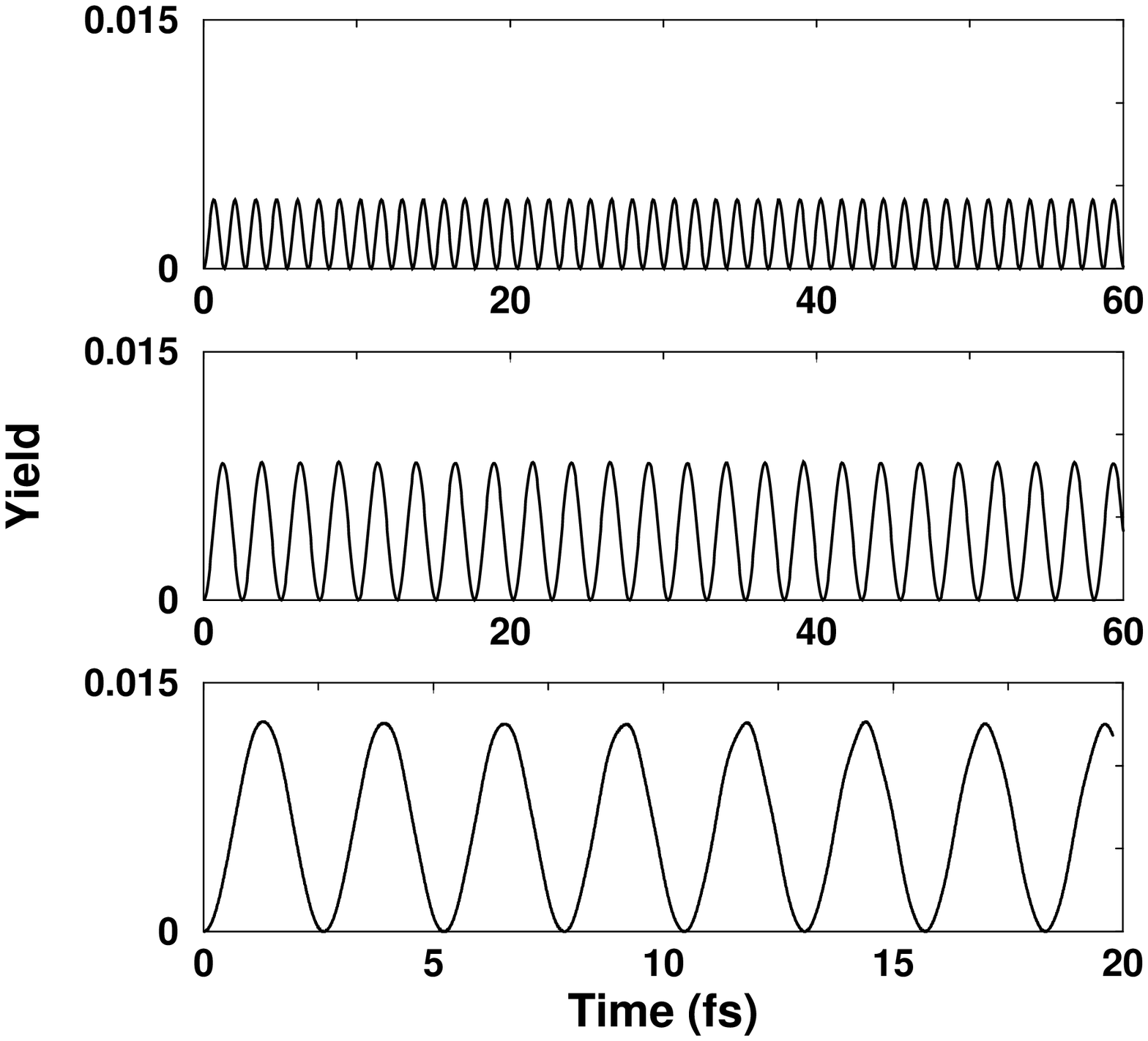}
\hspace*{0.6cm}\includegraphics[width=7cm,height=6.2cm]{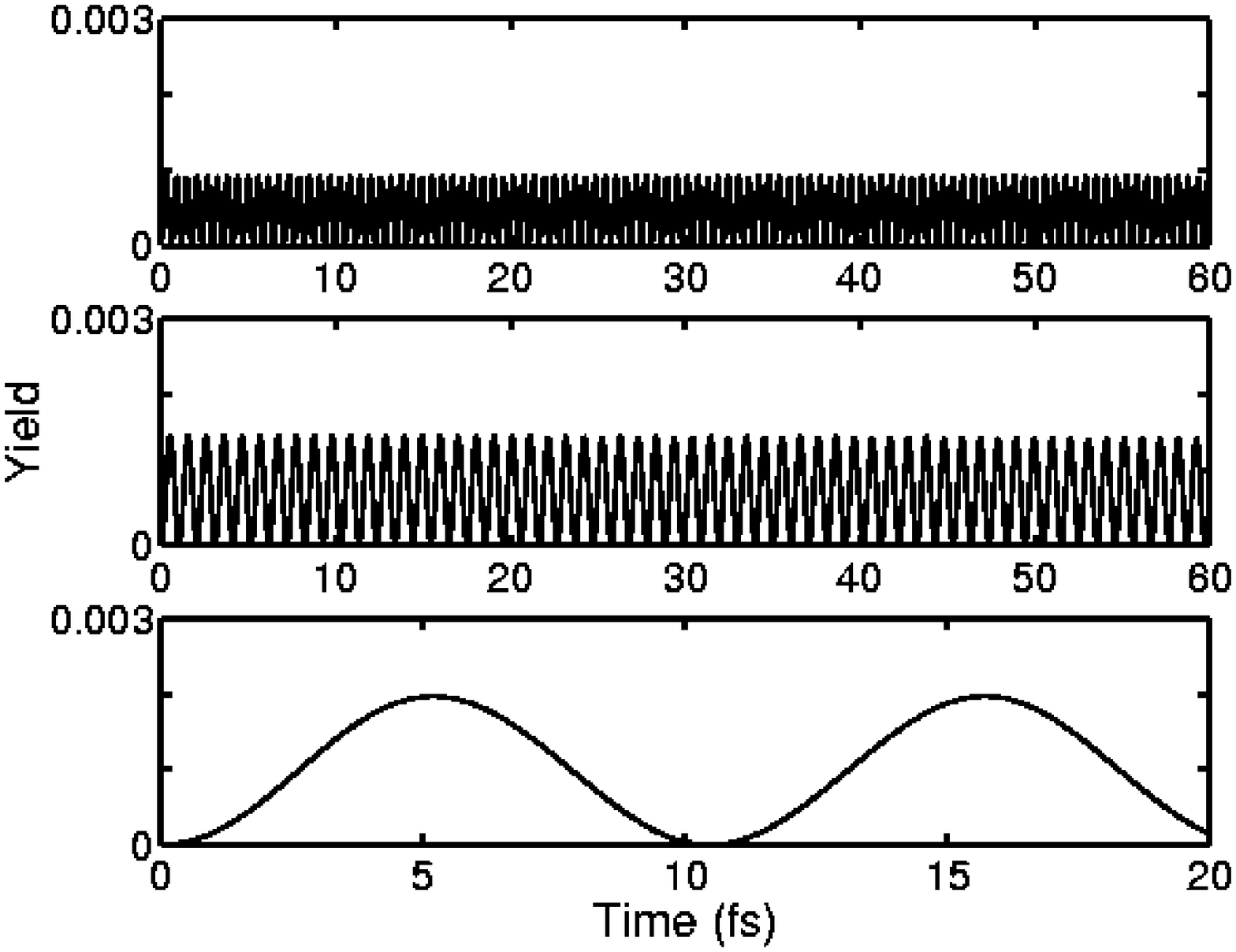}
\hspace*{0.2cm}\includegraphics[width=7.1cm,height=6.2cm]{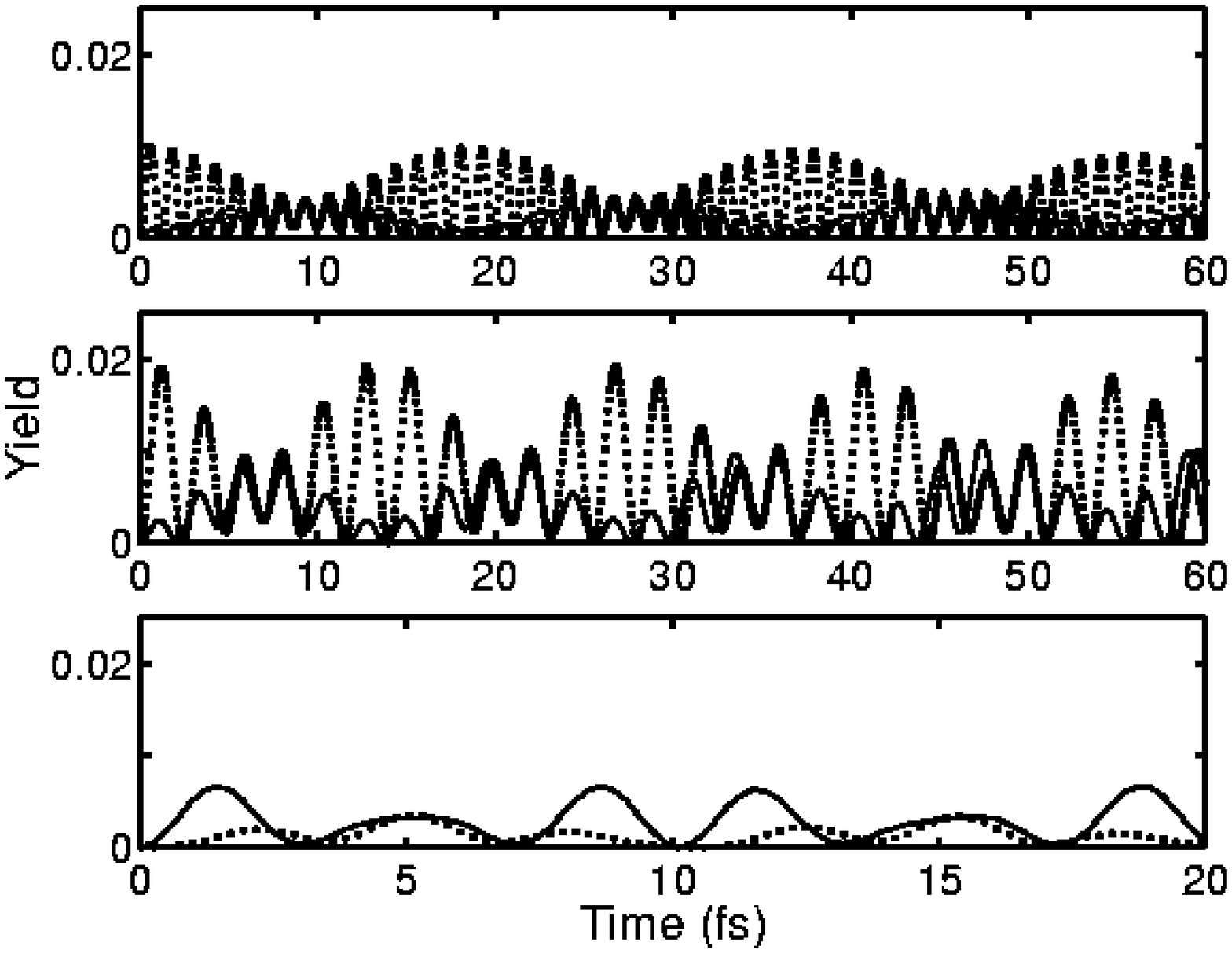}
\hspace*{0.1cm}\includegraphics[width=7cm,height=6.2cm]{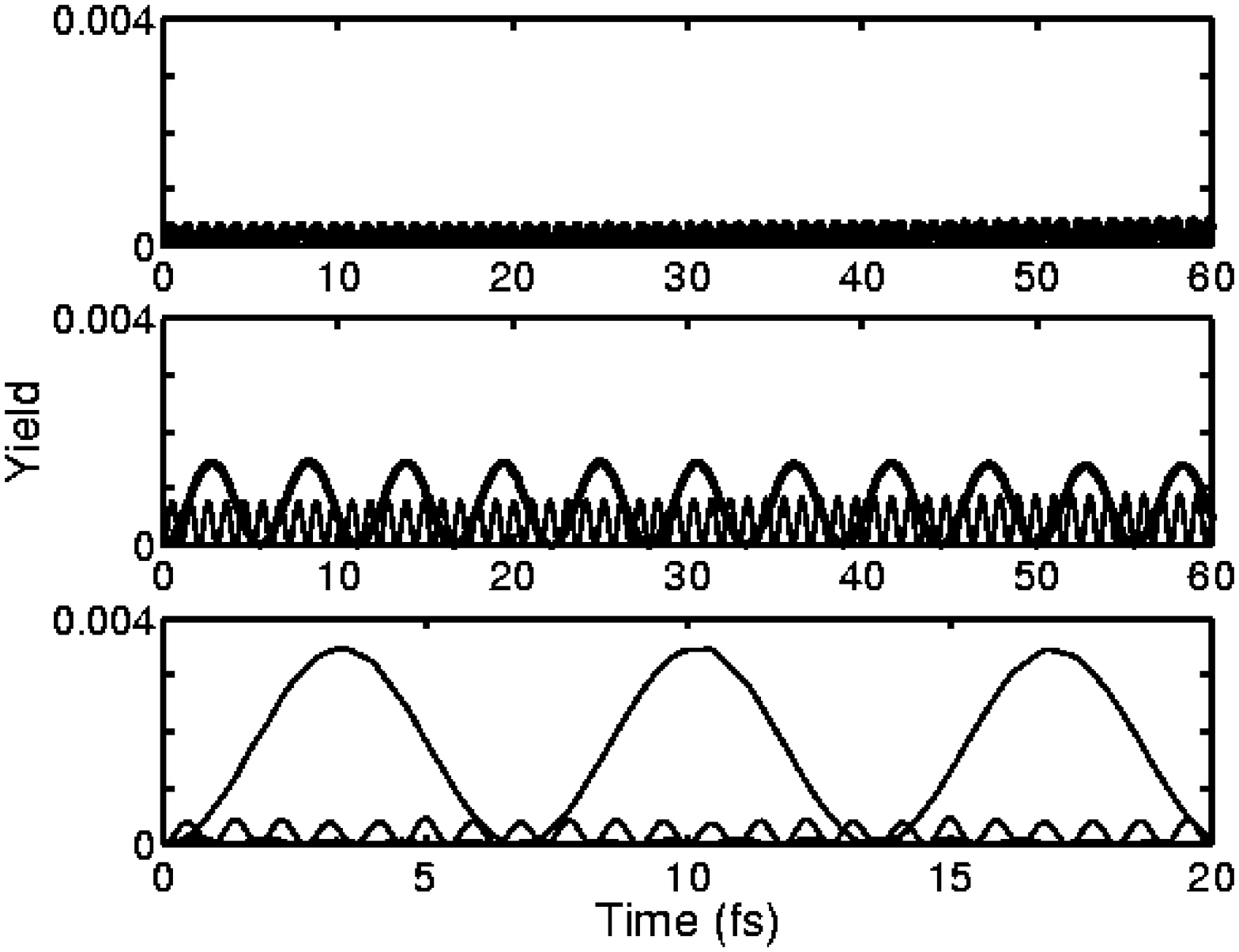}
\end{center}
\noindent
FIG. 5.{\it Yields in the singlet and triplet channels within the PPP
Hamiltonian. In all four cases the top panel corresponds to pair of ethylenes,
the middle panel to pairs of butadienes, and the bottom panel to
pairs of hexatrienes. Top left: singlet channel, $t_{\perp}=0.1{\rm eV},
~X_{\perp}=0$. Top right: triplet channel, $t_{\perp}=0.1{\rm eV},
~X_{\perp}=0$. Bottom left: singlet channel, $t_{\perp}=0.1{\rm eV},
~X_{\perp}=0.1 {\rm eV}$. Bottom right: triplet channel $t_{\perp}=
0.1{\rm eV},~X_{\perp}=0.1 {\rm eV}$. Evolution in case of hexatrienes 
is tracked for 20 fs while in other cases, the evolution is tracked for 
60 fs. Significant yields in singlet channel occurs only for final states 
$|(1^1A_g^+)_1(1^1B_u^-)_2 \rangle$ and $|(1^1B_u^-)_1(1^1A_g^+)_2\rangle$, 
between which the yields are identical for $X_\perp = 0$ but different for 
$X_\perp \ne 0$. Similar asymmetry is observed also in the triplet 
channels for $X_\perp \ne 0$.}
~\\
~\\
\noindent The differences from the H\"uckel model results are the following. 
First, the yields $I_{mn}(t)$ in both the singlet and triplet channels are 
considerably reduced in the PPP model.  Second, 
the 1$^1B_u^-$
yield is now substantially higher than that of the 1$^3B_u^+$ in all cases. 
Finally, the observed 
higher yield of the singlet exciton is true for both $X_{\perp}$ = 0 and 
$X_{\perp} \neq$ 0. 
This is in contradiction to the results obtained in references
\cite{Shuai00a,shuai2}, 
which ignore the energy difference between the 
1$^1B_u^-$ and the 1$^3B_u^+$. 
The only consequence of nonzero $X_{\perp}$ is 
the asymmetry between the yields of $(1^1A_g)_1(1^1B_u)_2$ and $(1^1A_g)_2
(1^1B_u)_1$ in the singlet channels, and a similar asymmetry in the triplet 
channels.
The overall conclusion that emerges from the results of the plots in
Figs. 5 is that nonzero electron-electron interactions substantially
enhances $\eta$.

\subsection{Finite size effects and their origin}

In what follows we will take the ground state energy E(1$^1A_g$) to be zero. 
In the infinite chain limit the sum total of the energies of the two 
oppositely charged polarons, E($P^+$) + E($P^-$), must be equal to the lowest 
singlet continuum band state in the neutral chain, independent of Coulomb 
interactions. Within the rigid band simple Hubbard model 
($V_{ij} = \epsilon_i$ = 0 in Eq.~\ref{PPP})
as $H_{intra}$, this implies that 
in the long chain limit E($P^+$) + E($P^-$) = E(1$^1B_u$). For nonzero 
intersite Coulomb interactions that are large enough to create an excitonic
energy spectrum, E($P^+$) + E($P^-$) = E(n$^1B_u$) in the long chain limit. 
Neither of these equalities 
are obeyed in short chains. In Table 1, we have listed 
E($P^+$) + E($P^-$) and E(1$^1B_u$) for the simple Hubbard model with 
$t_{ij} = t(1\pm\delta)$, $t$ = 1, and
$\delta$ = 0.2 for many different $U$, for N = 6. 
All quantities are in units of $t$.
We have also included here E(n$^3B_u$), defined in section 4.  In all cases 
E(1$^1B_u$) is significantly larger than E($P^+$) + E($P^-$), with the 
difference increasing with $U$. For nonzero intersite Coulomb interactions, 
E(n$^1B_u$) is similarly significantly higher than E($P^+$) + E($P^-$), 
as shown in Table 2 for N = 6, although the difference in energy decreases 
with increasing interaction strength,
due to localization. For sufficiently large Coulomb interactions, E(n$^3B_u$)
occurs below E($P^+$) + E($P^-$) for N = 6 within the extended Hubbard model. 
~\\
~\\
\noindent Table 1: {\it N = 6 energetics within the simple Hubbard model.
All energies are in units of t.}\\

\begin{center} \begin{tabular}{|c|c|c|c|}\hline
& & &  \\
\hspace{0.5cm}$U $\hspace{0.5cm}&
\hspace{0.5cm}$E(P^{+}) + E(P^{-})$ \hspace{0.5cm}&
\hspace{0.5cm}$E(1^1B_u)$\hspace{0.5cm}&\hspace{0.5cm}$E(n^3B_u)$\hspace{0.5cm}\\
\hspace{0.5cm}\hspace{0.5cm}&\hspace{0.5cm}\hspace{0.5cm}&
\hspace{0.5cm}\hspace{0.5cm}&
\hspace{0.5cm}\hspace{0.5cm}\\
& & &\\
\hline
& & &\\
2 &1.75  &1.99 & 2.94\\
 &  & &\\
4 &2.70  &3.05 &3.58 \\
 &  & &\\
6 &4.08 &4.48 & 4.84\\
 &  & &\\
8 &5.70  &6.12& 6.40 \\
 &  & &\\
10 &7.45  &7.89&8.12 \\
 &  & &\\
12 &9.28  &9.72 &9.92 \\
 &  & &\\
20 & 16.92 &17.38& 17.50  \\
 &  & &\\
\hline
\end{tabular} \end{center}
~\\
\noindent Table 2: {\it N = 6 energetics within the extended Hubbard model.
All energies are in units of t.}\\
\begin{center} \begin{tabular}{|c|c|c|c|c|}\hline
& & & &\\
\hspace{0.3cm}$U $\hspace{0.3cm}& \hspace{0.3cm}$V $\hspace{0.3cm}&
\hspace{0.3cm}$E(P^{+}) + E(P^{-})$ \hspace{0.3cm}&
\hspace{0.3cm}$E(n^1B_u)$\hspace{0.3cm}& \hspace{0.3cm}$E(n^3B_u)$\hspace{0.3cm}\\
\hspace{0.3cm}\hspace{0.3cm}&\hspace{0.3cm}\hspace{0.3cm} &
\hspace{0.3cm}\hspace{0.3cm}&\hspace{0.3cm}\hspace{0.3cm}&\hspace{0.3cm}
\hspace{0.3cm}\\
& & & &\\
\hline
& & & &\\
10 &3 & 7.52 & 9.05  & 7.67 \\
& & & &\\
10 &4& 7.32 & 8.98 &  7.06\\
 & & & &\\
12&4 & 9.26 & 10.66 & 8.77\\
 & & & &\\
20 &6 & 16.90 & 17.96 & 14.47\\
 & & & &\\
30 &10  &26.65 & 27.51 & 20.38 \\
 & & & &\\
\hline
\end{tabular} \end{center}
\vskip 1pc
The reason for this particular finite size effect 
is as follows. In the 
charged $P^{\pm}$ chains, there occurs a single carrier, a vacancy or a 
double occupancy within Hubbard and extended Hubbard models. 
Even for large Coulomb interactions, this carrier can be delocalized
over the entire chain (see Fig.~6). In contrast, in the 1$^1B_u$ or 
the n$^1B_u$ (and also in the n$^3B_u$) both the vacancy and the 
double occupancy are present, 
and hence the overall space left for the delocalization of any one carrier 
is considerably smaller. This reduced delocalization in the neutral chain 
increases the energies of the $^1B_u$ states 
(and also of the CT triplet state, n$^3B_u$)\cite{shuai97}. 
Furthermore, the ``squeezed'' nature of the wavefunctions of the 
excited states of neutral chains implies that the matrix elements of the 
type $\langle P^+P^-|H_{inter}|1^1A_gj^1B_u \rangle$ (and the corresponding 
matrix elements in the triplet channels) are also modified strongly in short 
chains. These finite size effects are larger in the higher energy states than 
in the lowest excitons (since only higher energy CT states have delocalized 
character in the long chain limit), and hence it is to be expected that the 
calculated yields of the higher energy singlet and triplet excitations for 
{\it realistic} Coulomb parameters in short chains may not be representative 
of the results expected for long chains. In order to understand long chain 
behavior, we will have to minimize the relative difference in the characters
of the neutral excited states and the charged polaron states. Since it is not 
possible to enhance the delocalizations of the double occupancy and
the vacancy of the 1$^1B_u$ and the n$^3B_u$, we will go to the opposite limit
of very strong Coulomb correlations, where the 1$^1B_u$, the n$^3B_u$ and
the $P^{\pm}$ are all nearly {\it equally localized}. 
As shown previously in the context of optical nonlinearity, the strong Coulomb
interaction limit for short chains can mimic the behavior of long chains with
intermediate Coulomb interactions \cite{DGuo}.
\begin{figure}
\begin{center}
{\includegraphics[height=6cm,width=5cm]{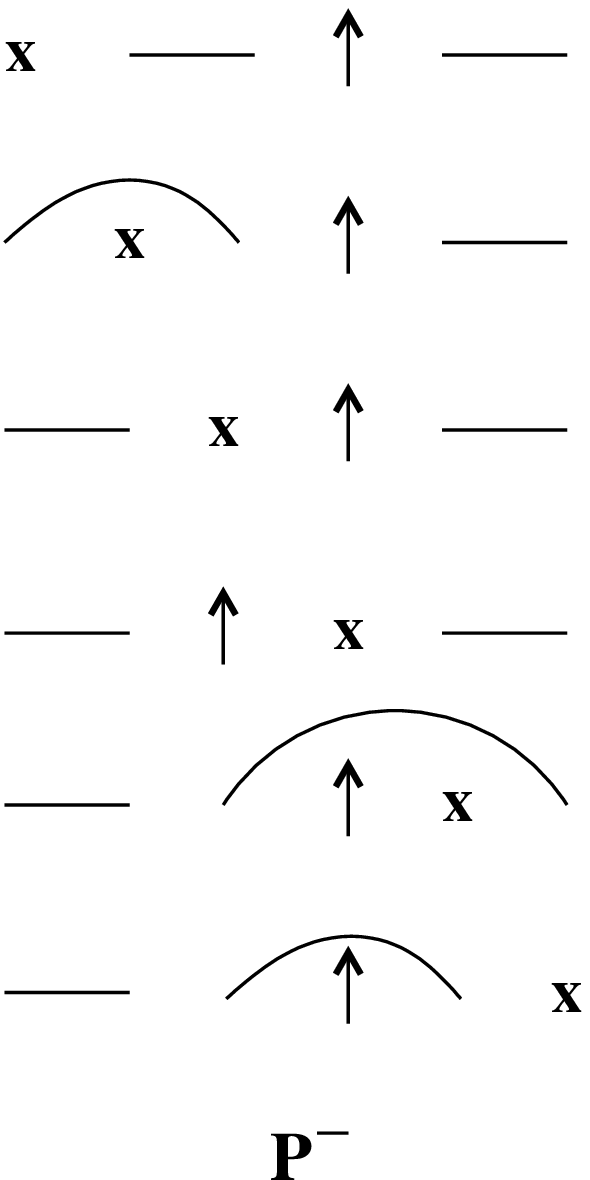}}
\end{center}
FIG. 6.: {\it A series of valence bond diagrams 
in which there 
occurs
delocalization of the doubly occupied site (denoted by $x$) in polaron $P^-$ from
left to right. The up arrow corresponds to an unpaired electron and a line between
two sites denotes singlet spin-pairing of the singly occupied orbitals at the sites.
In case of polaron $P^+$, the site with double occupancy ($x$) is
replaced by an empty orbital. 
In excited states containing both the double
occupancy and the vacancy in short chains, delocalization of each is 
considerably reduced, as they cannot pass one another without first going
through a virtual state with mutual annihilation.}
\end{figure}

\subsection{Hubbard model simulations}

We consider $H_{intra}$ as the  
simple Hubbard model ($V_{ij} = \epsilon_i = 0$ in Eq.~\ref{PPP})
with $t_{ij} = t(1\pm\delta)$, $t$ = 1 and
$\delta = 0.2$. For $H_{inter}$ we choose the $V_{ij} = X_{\perp}$ = 0
limit of Eq.~\ref{inter}, and $t_{\perp}$ =0.1. 
In what follows, we will no longer discuss the oscillatory behavior of
$I_{mn}(t)$, but will instead present the 
total yields $Y_{mn}$, obtained by integrating 
$I_{mn}(t)$ over the total duration of time evolution.
As discussed in the previous subsection, within the simple Hubbard model 
E(1$^1B_u) >$ E($P^+) +$ E($P^-$) and we do not expect any significant yield 
of $^1B_u$ states higher than the 1$^1B_u$. On the other hand,
with increasing $U$ the triplet 
energy difference $\Delta E(1^3B_u)$ = E($P^+$) + E($P^-$) -- E(1$^3B_u$) 
increases rapidly, while based on the discussions in section 4 we expect 
$\Delta E(n^3B_u)$ = E($P^+$) + E($P^-$) - E(n$^3B_u$) to decrease. 
Simultaneously, there occur many different triplet states below  
E($P^+$) + E($P^-$). It is of interest then to evaluate the yields of all
these triplet states and to determine whether or not the results based
on the PPP Hamiltonian survive for very strong Coulomb interactions.
As indicated below, by simultaneously monitoring the $Y_{mn}$, the energy 
difference between the final and initial states, and the corresponding matrix 
elements, we are able to obtain a useful mechanistic viewpoint of the e-h 
recombination.

In Figs. 7(a) and (b) we have summarized our results for the singlet channel. 
In spite of a thorough search, we did not find significant yield for any 
$^1B_u$ state other than the 1$^1B_u$. The yield for the 1$^1B_u$ decreases 
with $U$ for both N = 4 and 6, as shown in Fig.~7(a).
We have also evaluated the energy difference
$\Delta E(1^1B_u) = E(P^+) + E(P^-) - E(1^1B_u)$ as a function of $U$.
As shown in Fig.~7(b), $|\Delta E(1^1B_u)|$
which are in units of t
increases 
and the matrix element $\langle P^+P^-|H_{inter}|1^1A_gj^1B_u \rangle$ 
decreases with increasing $U$, suggesting that the yield scales as 
$\langle P^+P^-|H_{inter}|1^1A_gj^1B_u \rangle/|\Delta E(1^1B_u)|$, as 
would be true in a tunneling process. We have confirmed this scaling behavior
based on our data.
Interestingly, the matrix elements for 
N = 4 and 6 are nearly the same, and the higher yield in the longer chain is 
a simple consequence of the smaller $|\Delta E(1^1B_u)|$. 
\begin{figure}
\begin{center}
{\includegraphics[width=7cm,height=7cm]{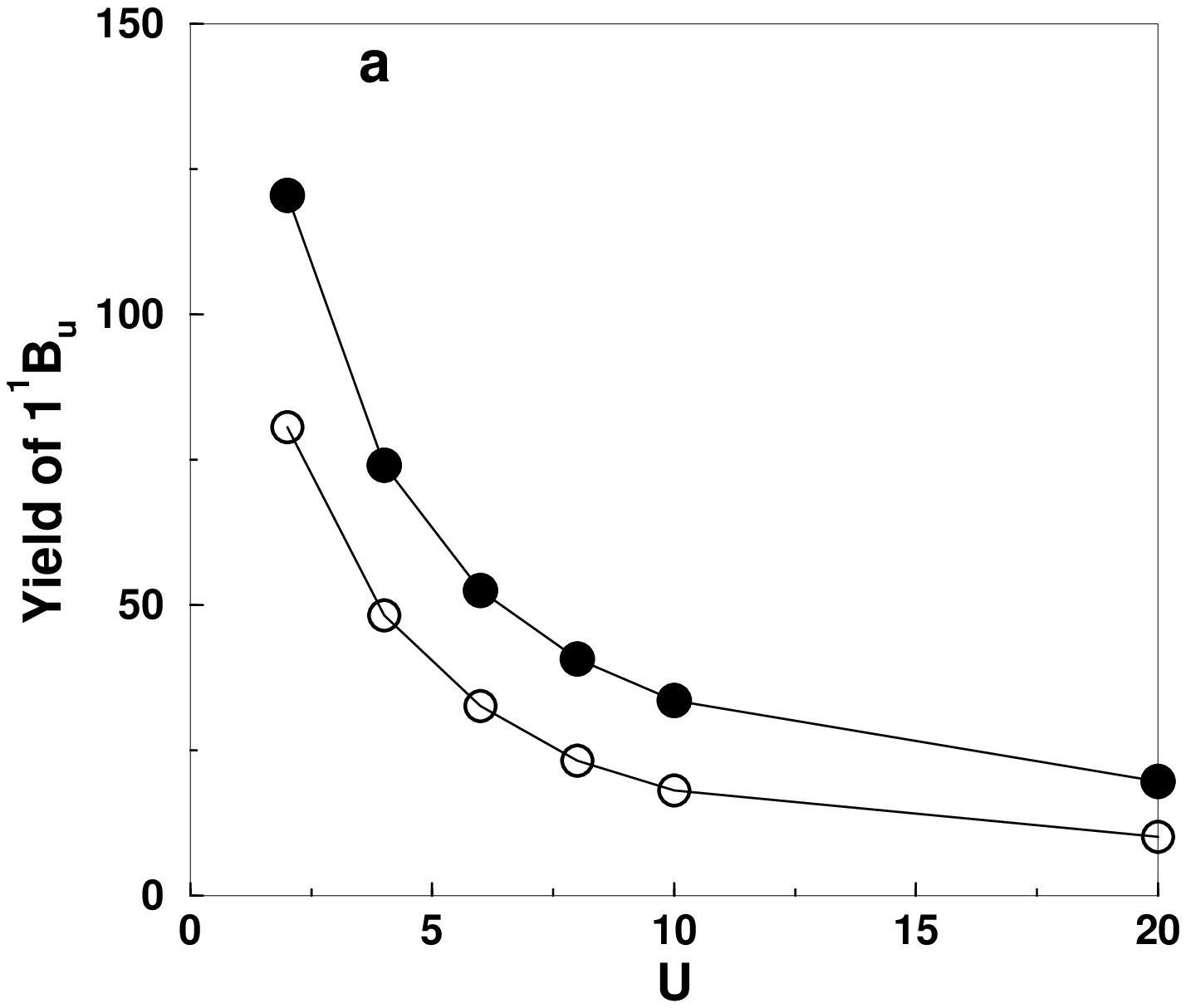}} \hfill
{\includegraphics[width=7cm,height=7cm]{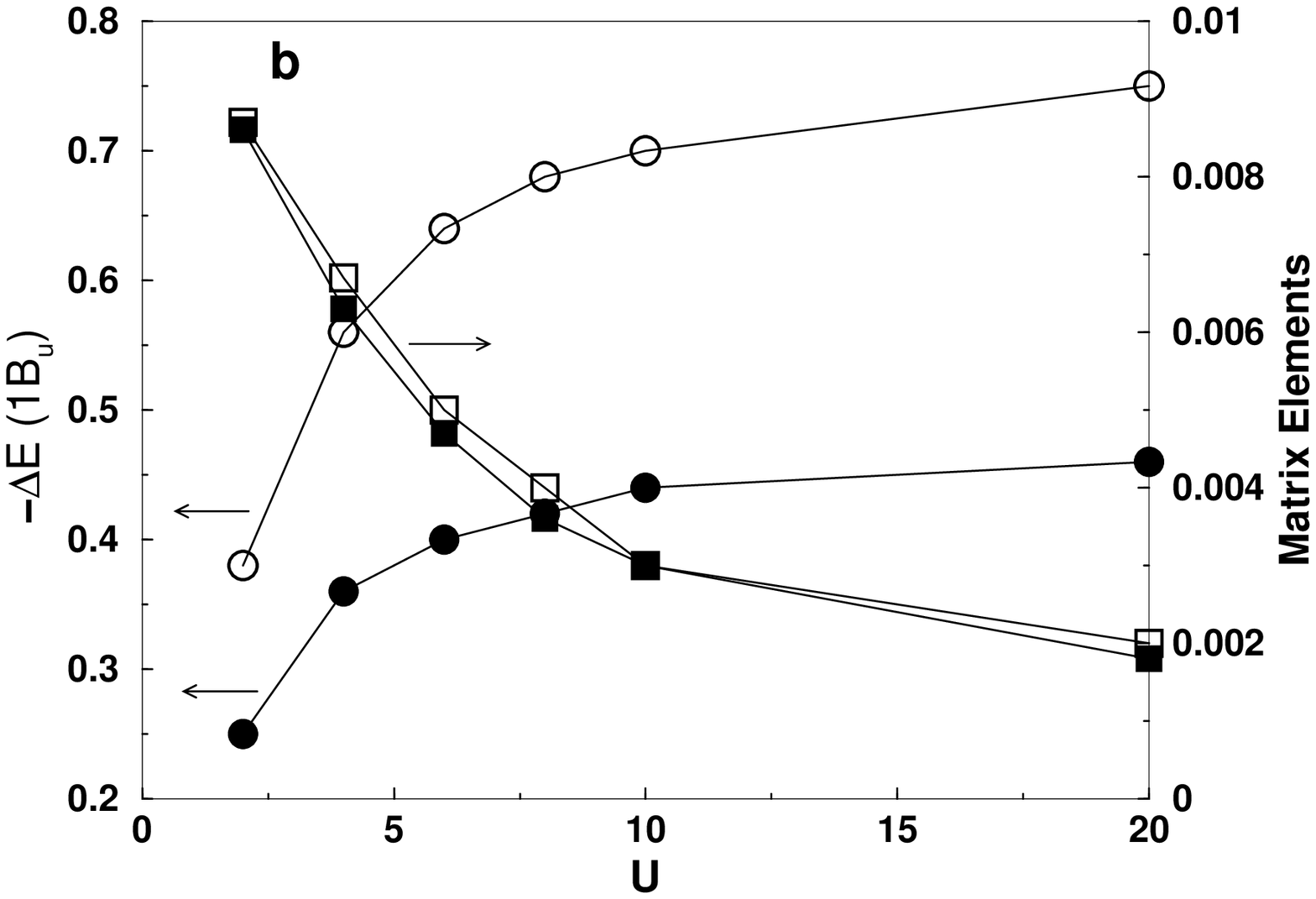}}
\end{center}
FIG. 7. {\it (a) Yield of $1^1B_u$ as 
a function of $U$ for a pair of butadienes (open circles)
and for a pair of hexatrienes (filled circles).
(b) Energy difference $\Delta E(1^1B_u)$ (see text)
and $\langle P^+P^-|H_{inter}|1^1A_g1^1B_u \rangle$
as functions of $U$ for a pair of
butadienes (open circles and open squares respectively) 
and for a pair of hexatrienes
(filled circles and filled squares respectively).} 
\end{figure}

\vskip 1pc
As expected, the behavior in the triplet channel is more complex. 
First of all, no $^3B_u$ state other than the 1$^3B_u$ and the n$^3B_u$ 
are generated in significant amounts, although several triplet states are 
found below the 1$^1B_u$ state for large $U$. This may be an artifact of the 
symmetry imposed by us on the two-chain model system (see section 7). 
More importantly, with increasing $U$, dominant triplet yield switches
from the 1$^3B_u$ state to the n$^3B_u$ state, as seen in Fig.~8 (a)
(n = 5 and 7 in butadiene and hexatriene, respectively).
In Fig.~8(b) we have shown the behavior of $|\Delta E(1^3B_u)|$ and 
$|\Delta E(n^1B_u)|$
(where the energy differences in units of t are again
with respect to $E(P^+) + E(P^-)$),
as well 
as the matrix elements $\langle P^+P^-|H_{inter}|1^1A_g1^3B_u \rangle$ and 
$\langle P^+P^-|H_{inter}|1^1A_gn^3B_u \rangle$ for the case of N = 4 (the 
behavior of these quantities for N = 6 are identical). The rapid increase of 
$\Delta E(1^3B_u)$ and the decrease of $\langle P^+P^-|H_{inter}|1^1A_g1^3
B_u \rangle$, shown in Fig.~8(b), explain the rapid decrease in the 
1$^3B_u$ yield seen in 
Fig.~8(a).  E(n$^3B_u$) is higher than E($P^+$) + E($P^-$) for all values 
of $U$ (which as pointed out in the above, is a finite size effect) and the 
$|\Delta E(n^3B_u)|$ decreases very slowly with increasing $U$. The 
matrix element $\langle P^+P^-|H_{inter}|1^1A_gn^3B_u \rangle$ remains almost 
a constant over the complete range of $U$ we have studied. Thus the initial
increase in the yield of the n$^3B_u$ followed by its saturation is expected 
from the behavior of $\langle P^+P^-|H_{inter}|1^1A_gn^3B_u \rangle/|\Delta 
E(n^1B_u)|$.  Interestingly, $\langle P^+P^-|H_{inter}|1^1A_g1^3B_u \rangle$ 
continues to be 
larger than $\langle P^+P^-|H_{inter}|1^1A_gn^3B_u \rangle$ even in the region 
where the yield of the n$^3B_u$ is higher, indicating once again that both the 
matrix element and the energy difference between the initial and final states 
determine the yield in any given channel. Taken together, the results of 
Figs.~8 also suggest that in the triplet channels, matrix elements favor 
higher yield for the 1$^3B_u$, but energetics favor higher yield for the 
n$^3B_u$.   

\begin{figure}
\begin{center}
{\includegraphics[width=7cm,height=7cm]{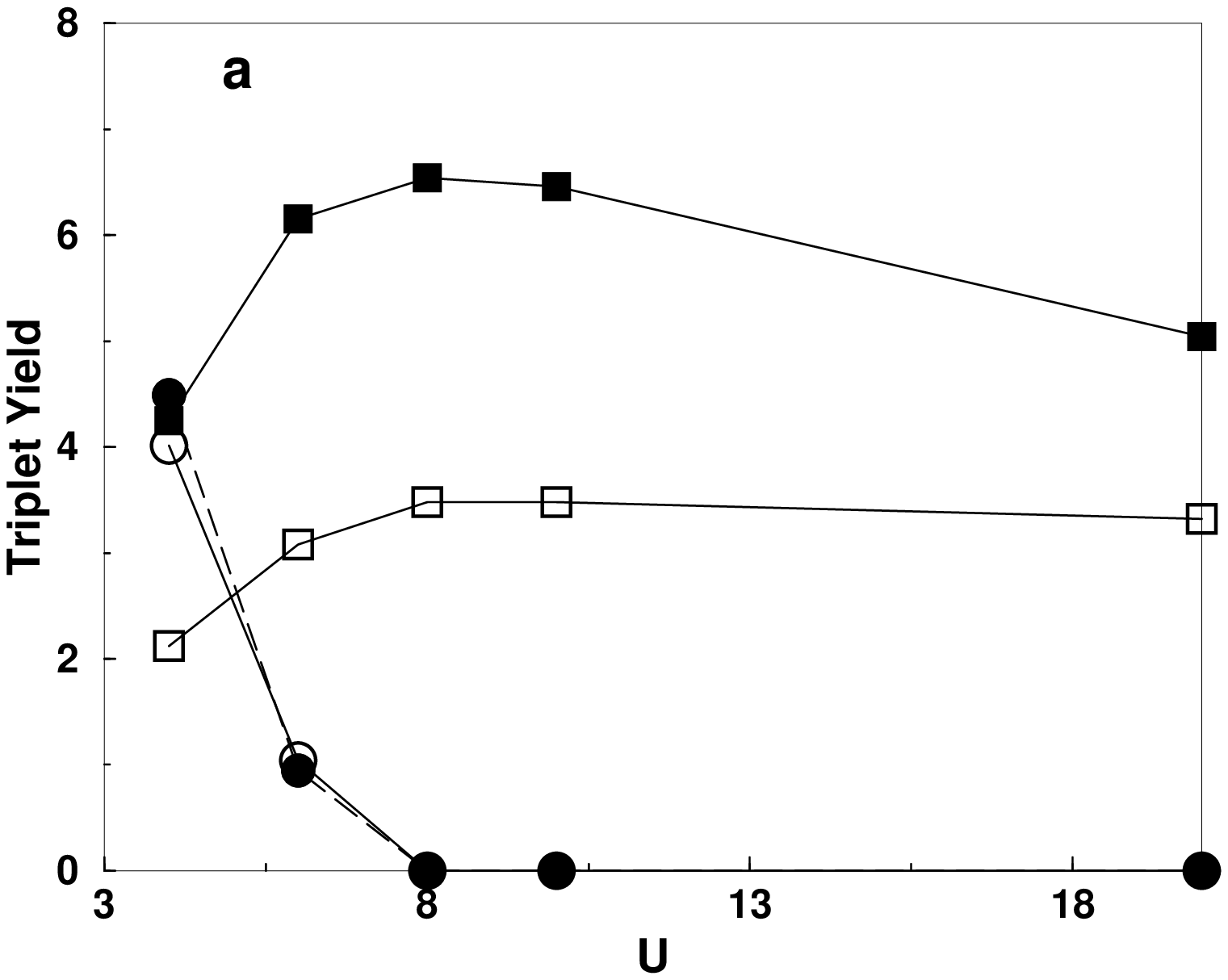}}
\hfill {\includegraphics[width=7cm,height=7cm]{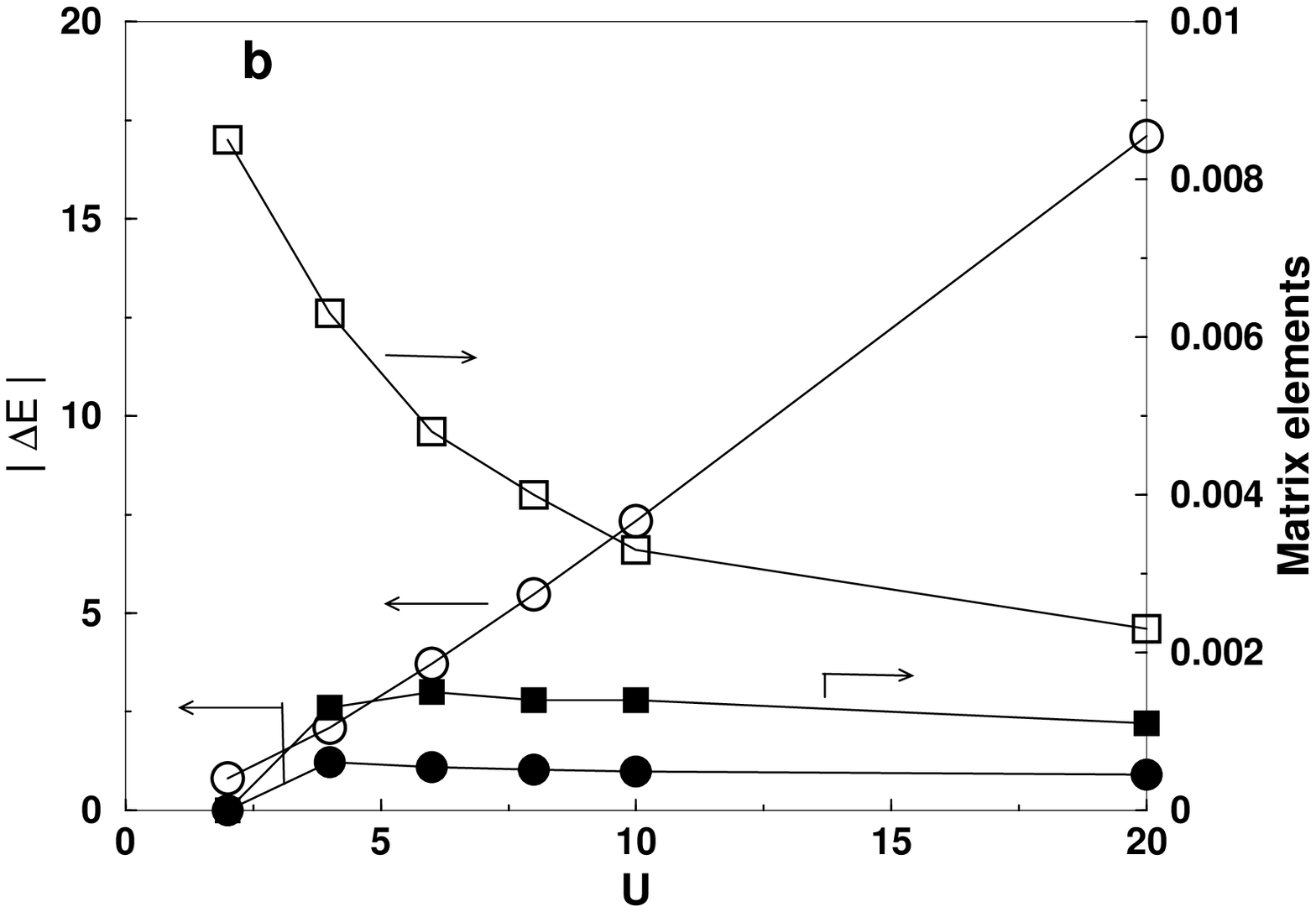}}
\end{center}
FIG. 8. {\it (a) Yields for the $1^3B_u$ and 
$n^3B_u$ as
a function of $U$, for a pair of butadienes (open circles and open squares
respectively) and for a pair of hexatrienes (filled circles and filled squares
respectively).
(b) $|\Delta E(j^3B_u)|$ and $\langle P^+P^-|H_{inter}|1^1A_gj^3B_u\rangle$
versus $U$ for $j$ = 1 (open circles and open squares) and $j$ = n (filled 
circles and filled squares) for a pair of
butadienes.
Unlike in Fig.~7, we have plotted here the absolute energy differences, as
the $n^3B_u$ can occur both above and below the $|P^+P^-\rangle$ due to
finite size effects.}  
\end{figure}

The time-dependent perturbation theory \cite{Merzbacher}
shows that the transition probability
to a specified excited state $|k>$ from an initial state $|i>$ is given by 
\begin{eqnarray}
P_{i \to k} = \left |{{\left<k|H^{'}|i\right>}\over{E_k-E_i}}\right |^2
\end{eqnarray}
In this spirit, we compute the transition probability to all the dominant
singlets as well as triplets. From these transition probabilities, we
compute $\eta_{TP}$ for various values of the Hubbard parameter $U$. We also 
compute $\eta_D$ from the yields to all the singlet and triplet states obtained
from our dynamical simulations. These two are shown as a function of $U$
in Fig. 9 for a pair of hexatrienes. Similar behavior is also seen for
a pair of butadienes. We note that the two different
approaches give qualitatively the same behavior. This clearly vindicates 
our focus on the matrix elements of the interchain interactions and
the energy differences between the initial and final states.
\begin{figure}
\begin{center}
{\includegraphics[width=10cm,height=9.0cm]{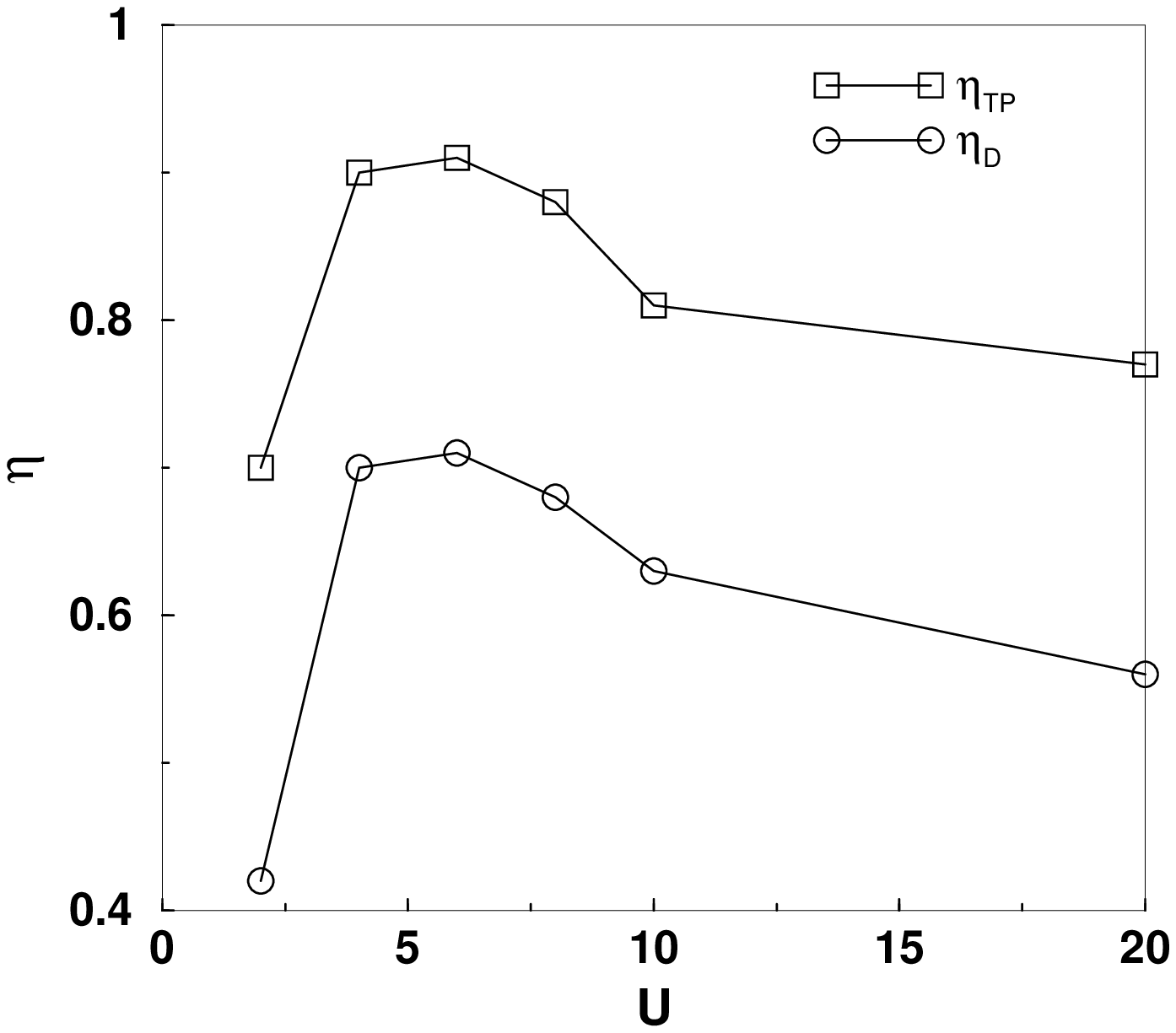}}
\end{center}
\noindent {FIG. 9.{ \it Variation of $\eta$ as a function of $U$ for 
a pair of hexatrienes.}}
\end{figure}

\vskip 1pc
Although the above finite size calculations by themselves have limited scope,
we believe that they are quite instructive. The behavior in the triplet channel 
clearly shows the bifurcation of the reaction paths, with the relative weights 
of the two paths being a strong function of the Coulomb parameter. 
The energy difference factor is large for the 1$^3B_u$ (which has the larger 
matrix element with the initial reactant state), while the matrix element is 
smaller for n$^3B_u$ and the energy difference is smaller. The 1$^1B_u$ has 
both a large matrix element (as the 1$^3B_u$) 
as well as a small energy
difference (as the n$^3B_u$), and hence its yield is larger than the overall
triplet yield which is the sum total of the yields of the 1$^3B_u$ and the
n$^3B_u$. We believe that this particular result continues to be valid 
qualitatively for long chains
with realistic Hubbard $U$. 

\subsection{Simulations within the extended Hubbard model}

The simple Hubbard model does not lead to exciton formation and the singlet 
yield is limited to the 1$^1B_u$. In order to see the bifurcation of the 
e-h reaction path in the singlet channel, one therefore has to work with
$H_{intra}$ corresponding to
extended Hubbard models which support an excitonic electron structure. For 
moderate Coulomb interactions, as in the PPP model, the bifurcations are 
washed out due to the finite size effects discussed in section 6.3. We 
perform our calculations again for very strong Coulomb interactions
in $H_{intra}$, 
where finite size effects are minimized due to extreme localization, both 
in the charged and neutral systems. Furthermore, we restrict the intersite 
Coulomb interactions to nearest neighbors only, to minimize the particle-hole 
separation in the 1$^1B_u$ exciton state and generate very strongly bound 
exciton. This procedure ensures that there exist distinct delocalized CT 
states above the exciton even in short chains (see section 4). We have again
chosen $X_{\perp}$ = 0 and $t_{\perp}$ =0.1 in $H_{inter}$. 
In order to be consistent with nonzero intrachain intersite Coulomb 
interaction, we have
now included interchain 
$V$ = 10\% of intrachain $V$.

The results of our calculations are shown in Table 3, where we have listed 
the yields of the two dominant singlet (1$^1B_u$ and n$^1B_u$) and 
dominant
triplet (1$^3B_u$ and n$^3B_u$) states, the energy differences $\Delta E(j^1B_u)$ 
and $\Delta E(j^3B_u)$ ($j$ = 1 and n), defined as before with respect
to E($P^+$) + E($P^-)$, and the relevant matrix elements of $H_{inter}$ 
between the initial and various final states. Several conclusions emerge 
from these data. 

(1) For such large Coulomb interactions the n$^3B_u$ is 
(for moderate $U$ = 10, $V$ = 3 and 4) energetically close to 
the initial state even as the n$^1B_u$ is considerably higher in energy 
(we have already argued that the latter is a finite size effect
\cite{Barford}). As in the 
previous subsection, the bifurcation in the triplet channel leads to a very 
high yield of the n$^3B_u$. Differently from the previous case though, the 
exciton character of the 1$^1B_u$ ensures a large $\Delta E(1^1B_u)$ 
in the present case,
and hence a small yield of the 1$^1B_u$. Thus in this narrow regime of 
Coulomb interactions, the triplet yield dominates over the singlet yield. 
This is an artifact of our restriction to short chains, as explained below. 

(2) As the Coulomb interactions are increased further, the bifurcation 
in the singlet channel reaction path sets in and in this case
the n$^1B_u$ yield dominates over that of the 1$^1B_u$. 
Indeed in this region the $\Delta E(n^3B_u)$ is moderately large once again even as $\Delta E(n^1B_u)$ is small (though still negative). Thus in the limit of 
very large Coulomb interactions the overall singlet yield again dominates 
over the triplet yield, as within the PPP model, with the difference
that here in both the spin channels the higher energy state dominates over the
corresponding lower energy state. 

(3) Matrix elements of $H_{inter}$ between initial and final states
are not independent of the energy difference between them, -- smaller the 
energy difference larger is the matrix element. This makes understanding the 
finite size effects extremely important, since in short chains where the 
higher energy singlet and triplet states are much too high in energy, 
matrix elements leading to these states are {\it simultaneously} small, 
thereby reducing the overall yields to these states. In both singlet and 
triplet channels, we expect the bifurcations of the reaction paths to play 
important roles in the long chain limit. 

(4) Finally, we note that even as the energy differences between the 
initial polaron-pair state and higher (lower) energy final states become
small (large), the relevant matrix elements continue to be large for the
lower states 1$^1B_u$ and 1$^3B_u$. We believe that this will continue to be 
true in the long
chain limit, the implication of which is that the matrix elements favor
the lower excitons, while the smaller energy difference favors the higher
energy CT states. This result is the same as that observed in the triplet
channel for the simple Hubbard model.
The true sum total yields in either spin channel is therefore
very difficult to calculate directly, and proper implications of the above
data should be sought.
~\\\\
Table 3: {\it The energy differences, matrix elements and yields
of singlet and triplet $B_u$ states for various values of $U$ and $V$ 
parameters of 
N=6 with $H_{intra}$ as the extended Hubbard model. 
For each set of $U$ and $V$, 
the first row
corresponds to S = 0 and the second row corresponds to
S = 1. The total singlet to triplet yield ratio is small at the 
top but high at the bottom of the table.
All energies are in units of t.} 
\begin{center} \begin{tabular}{|c|c|c|c|c|c|c|c|}\hline
& & & & & & & \\
\hspace{0.07cm}$U$\hspace{0.07cm}&\hspace{0.07cm}$V$\hspace{0.07cm}&
\hspace{0.1cm}$\Delta E(1B_u)$\hspace{0.1cm}&
\hspace{0.1cm}$\Delta E (nB_u)$\hspace{0.1cm}&
\hspace{0.1cm}$H^{\prime}_{P^+ \otimes P^-~;~ G\otimes 1B}$\hspace{0.1cm}
& \hspace{0.1cm}$H^{\prime}_{P^+ \otimes P^-~;~ G\otimes nB} $\hspace{0.1cm}
&\hspace{0.1cm} Y(1B) \hspace{0.1cm}&\hspace{0.1cm} Y(nB)\hspace{0.1cm} \\
\hspace{0.07cm}\hspace{0.07cm}&\hspace{0.07cm}\hspace{0.07cm} &
\hspace{0.1cm}\hspace{0.1cm} &\hspace{0.1cm}\hspace{0.1cm} &$(10^{-3})$&
$(10^{-3})$ & &\\
& & & & & & &\\
\hline
& & & & & & &\\
 & & 0.84 & -1.53  & 4.12& 0.35& 16.84& \\
10& 3& & & & & &\\
& & 6.93 & -0.14  & 3.21& 1.64& 0.18&77.36 \\
& & & & & & &\\
\hline
\hline
& & & & & & & \\
& & 1.83 & -1.65  & 3.49& 0.53 & 3.00& \\
10&4 & & & & & &\\
& & 6.64 & 0.27  & 3.39& 1.45 & 0.21&66.67 \\
& & & & & & & \\
\hline
\hline
& & & & & & & \\
& & 1.60 & -1.40  & 3.33& 0.54 & 4.27& \\
12&4 & & & & & &\\
& & 8.74 & 0.50  & 2.88& 1.41 & 0.11&17.88 \\
& & & & & & & \\
\hline
\hline
& & & & & & & \\
& & 3.13 & -1.06  & 2.10& 0.64 & 0.65& 0.87 \\
20&6 & & & & & &\\
& & 16.58 & 2.43  & 2.03& 0.92 & &0.43 \\
& & & & & & & \\
\hline
\hline
& & & & & & & \\
& & 6.86 & -0.86  & 1.53& 0.59 & 0.11& 0.76 \\
30&10 & & & & & &\\
& & 26.42 & 6.26  & 1.70& 0.68 & & \\
& & & & & & & \\
\hline
\end{tabular} \end{center}
~\\


\section{Discussion and conclusions}
\label{conclusion}

Although our calculations are for finite systems, the have the advantage of 
being exact. Based on our experience \cite{DGuo}, we believe that true long
chain behavior for realistic Coulomb interactions can be obtained by 
``grafting together'' the different pieces of information discussed in sections
6. The interpretation of the PPP model calculations is obvious:
although 
$|\langle 1^1A_g \cdot 1^1B_u |H_{inter}| P^+P^- \rangle_S| \sim
|\langle 1^1A_g \cdot 1^3B_u |H_{inter}| P^+P^- \rangle_T|$
with these parameters, the
proximity of the S = 0 final state to the initial polaron-pair state 
relative to the S = 1
final state favors the singlet, thereby leading to $\eta$ considerably
larger than 0.25 \cite{Tandon,Moushumi}. No significant yield to higher energy
states are obtained within the PPP model at our chain lengths, but as we have
shown, this is because the magnitudes of the matrix elements between initial 
and final states depend on the energy differences between them, and at these short
chain lengths, the relevant energy differences are too large. Thus the PPP model
results cannot be thought of as complete. The results in section 6.4 are very
instructive. The relatively large (small) energy difference between the 
1$^3B_u$ (n$^3B_u$) and the
initial state composed of the polaron pairs that occurs here for large Hubbard $U$
is exactly what is expected at long chain lengths for intermediate $U$. We expect
then that in long chains 
dominant singlet yield to the 1$^1B_u$ and dominant triplet yield to the n$^3B_u$.
The continuing higher yield of the singlet, even when compared to the total triplet
yield, is then significant. The reason for the relatively lower triplet yield can
be understood from Fig.~8(b), -- even as 
$|\langle 1^1A_g \cdot 1^3B_u |H_{inter}| P^+P^- \rangle_T|$
decreases with $U$, it 
remains larger than 
$|\langle 1^1A_g \cdot n^3B_u |H_{inter}| P^+P^- \rangle_T|$.
In contrast, the 1$^1B_u$ remains energetically proximate to $|P^+P^-\rangle_S$, 
and hence
the matrix element in the singlet channel also continues to be larger. The 
resultant large $\eta$ within simple Hubbard models is therefore expected even
in the long chain limit.

The behavior within the extended Hubbard model is only slightly more complex.
Once again, we believe that increasing Coulomb correlations at fixed N
is qualitatively equivalent to increasing N at fixed Coulomb correlations, since
both have the same effect on the order and proximities of the most relevant
energy states. We have not included the results for smaller $U$, $V$ in Table 2,
because at these parameters the behavior continues to be similar to that within the
PPP model, i.e., the yields are dominated by the lowest S = 0 and 1 excitons,
with $\eta >$ 0.25. With
stronger Coulomb interactions, in both the singlet and triplet channels
the yield shifts from the lowest excitons to the higher n$B_u$ states. 
Nevertheless, $\eta$ is predicted to be greater than 0.25 for most of the
parameter space.
Unfortunately,
we are unable to demonstrate the gradual shift from lower energy to higher energy
products as would occur upon increasing N at fixed Coulomb correlations: our
reaction products are either the lower energy states or the higher energy ones.
We believe that the
true long chain behavior is very likely a ``superposition'' of the calculated
results for moderate and strong Coulomb interactions, viz., larger matrix elements
with the lower states as final states favoring these, while proximities in
energy favoring the higher CT states. For systems with relatively small singlet 
exciton
binding energies, in the long chain limit we expect the yield in the singlet
channel to be dominated by the 1$^1B_u$, as in the simple Hubbard model. For
systems with large singlet exciton binding energy, it is conceivable that the
n$^1B_u$ and the n$^3B_u$ both make significant contributions. The relative
yield of the 1$^3B_u$ in these systems should be very small, and as a consequence
we expect again the overall $\eta$ to be large. To summarize then, what varying
chain lengths, Coulomb parameters and exciton binding energies do is redistribute
the overall yields within each spin channel between the lowest exciton and the higher
energy CT state. However, even after this redistribution $\eta$ remains greater than
0.25. Clearly, completely convincing proof of the above conjecture would require
long chain calculations that can treat low and high energy states as well
as the polaron-pair states of long chains with high precision. We are
pursuing density matrix 
renormalization group calculations currently to test the ideas posed in the above.

One apparent surprise in our results is the overall limitation to the 1$^3B_u$ and
the n$^3B_u$ as products in the triplet channel e-h recombination. This is
surprising, given that there exist so many covalent triplet states between the
1$^3B_u$ and the n$^3B_u$ even in short chains (Fig. 10). 
\begin{figure}
\begin{center}
{\includegraphics[width=10cm,height=7.5cm]{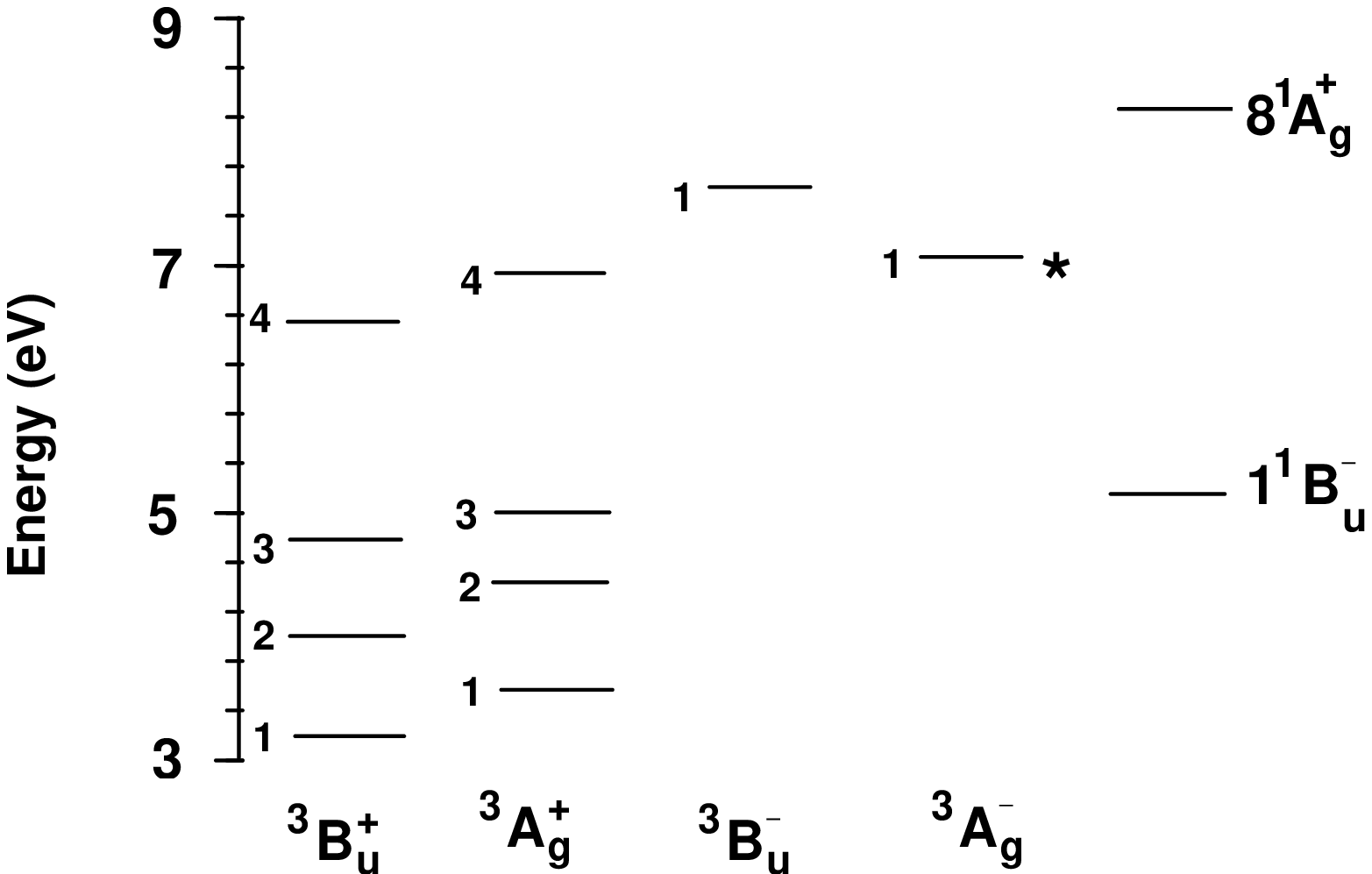}}
\end{center}
{FIG. 10.{ \it The triplet energy spectrum between the 1$^3{\rm B}_u^+$ and
1$^3{\rm A}_g^-$ in a N = 12 chain relative to the singlet ground state, within
the PPP-Ohno Hamiltonian.
Different symmetry subspaces are shown separately. State marked by asterisk
is dipole-coupled to the 1$^3{\rm B}_u^+$. The 1$^1{\rm B}_u^-$ and the
$m^1{\rm A}_g^+$ ($m$ = 8 in N = 12) are also included.}}
\end{figure}

Indeed, if these triplets were
generated, in principle $\eta$ could have reached values smaller than 0.25.
One reason these triplets are not generated within our calculations is because 
of the artificial mirror plane symmetry that we have imposed between the two
polyene chains, as a consequence of which even the charge-separated polaron-pair
states have $B_u^-$ and $B_u^+$ symmetries in singlet and triplet spin spaces,
respectively, and the interchain hop maintains these symmetries. All
low to intermediate energy triplets that belong to symmetry subspaces different
from the $^3B_u^+$ are thereby excluded. There do exist, however, $^3B_u^+$
states above the 1$^3B_u$ but below the n$^3B_u$, and even these are not generated
in significant amounts over a broad region of the parameter space. We are
currently pursuing exact calculations with all possible relative orientations
between the two molecular components. While these calculations will obviously
generate singlets and triplets belonging to all possible symmetry subspaces, we
believe that they will demonstrate the existence of approximate ``sum rules'',
i.e., the total yield in each spin channel remains nearly conserved, and that the
total yields are close to what we have already found in our current calculations. 

Finally, we address the issue of electron-phonon interactions, ignored in our
calculations. Electron-phonon interactions play a dominant role in
theories of e-h recombination in which intermolecular charge-transfer leads to
higher energy singlet and triplets only, and differences in cross-sections
arises from differences in structural relaxations to the lowest excitons
\cite{hong,bittner}. As we have pointed out elsewhere, these theories
ignore triplet states in between the high energy charge-transfer state and
the lowest triplet, and such triplets should definitely be involved in
nonradiative relaxation \cite{MDas}. Inclusion of these intermediate triplets
will enhance the triplet relaxation and wipe out all differences between
singlet and triplet relaxations that give the difference between singlet and
triplet yields within the above theories. The only way electron-phonon 
interactions can influence the singlet:triplet yield within our picture
is if these interactions change substantively the relative energies of the
most relevant states. It has been claimed by Conwell, for example, that 
the relaxed polaron energies E($P^+$)+E($P^-$) can be considerably below  
the E(n$^1B_u$) that is observed in nonlinear optical experiments 
\cite{Conwell}. This has, however, not been substantiated by any calculations,
and we are currently investigating this possibility.
\vskip 1pc
\section{Acknowledgments}

Work in Bangalore was supported by the CSIR, India and DST, India,
through /INT/US
(NSF-RP078)/2001. Work in Arizona was supported
by NSF-DMR-0101659, NSF-DMR-0406604 and NSF-INT-0138051. 
We are grateful to our experimental
colleagues Z.V. Vardeny and M. Wohlgenannt for numerous stimulating
discussions. S.M. acknowledges the hospitality of the Indian Institute of
Science, Bangalore, where this work was completed.

{}

\end{document}